\documentclass{article}
    \PassOptionsToPackage{numbers,sort,compress}{natbib}
\usepackage[]{new_arxiv}

\usepackage[utf8]{inputenc} %
\usepackage[T1]{fontenc}    %
\usepackage{hyperref}       %
\usepackage{url}            %
\usepackage{booktabs}       %
\usepackage{amsfonts}       %
\usepackage{nicefrac}       %
\usepackage{microtype}      %

\usepackage{times}
\usepackage{epsfig}
\usepackage{graphicx}
\usepackage{amsmath}
\usepackage{amsmath,amssymb,mathtools}

\usepackage{amsmath,amsfonts,bm}

\def\eqref#1{equation~\ref{#1}}

\def\1{\bm{1}}

\def\vv{{\bm{v}}}

\DeclareMathAlphabet{\mathsfit}{\encodingdefault}{\sfdefault}{m}{sl}
\SetMathAlphabet{\mathsfit}{bold}{\encodingdefault}{\sfdefault}{bx}{n}

\def\gL{{\mathcal{L}}}

\newcommand{\cut}[1]{}

\newcommand{\prep}{\mathbf{r}}

\DeclareMathOperator*{\argmin}{arg\,min}

\DeclareMathOperator{\sign}{sign}

\usepackage{url}
\usepackage{makecell} 
\usepackage{pdfpages}
\usepackage{adjustbox}
\usepackage{wrapfig}

\usepackage[utf8]{inputenc}

\usepackage{multirow}
\usepackage[utf8]{inputenc} %
\usepackage[T1]{fontenc}    %
\usepackage{booktabs} %
\usepackage{balance} %
\usepackage{amssymb}
\usepackage{subcaption}
\usepackage{array} 
\usepackage{multirow}
\usepackage[linesnumbered,ruled]{algorithm2e}
\usepackage{algpseudocode}
\usepackage{amsmath,amssymb,mathtools}
\usepackage{cuted}
\usepackage{amsthm,bm}
\usepackage{afterpage}
\usepackage{bbm}
\usepackage{color}
\usepackage{algorithm2e}
\usepackage{mathtools}

\makeatletter
\DeclareRobustCommand\onedot{\futurelet\@let@token\@onedot}
\def\@onedot{\ifx\@let@token.\else.\null\fi\xspace}

\def\eg{\emph{e.g}\onedot} 
\def\ie{\emph{i.e}\onedot} 
 
\def\etc{\emph{etc}\onedot} 
\def\wrt{w.r.t\onedot} 

\makeatother

\newcommand{\adv}{\mathrm{adv}}

\usepackage{xspace}
\usepackage{gensymb}

\SetCommentSty{mycommfont}

\newcommand{\reftbl}[1]{Table~\ref{tbl:#1}}

\newcommand{\shortrefsubsec}[1]{\S~\ref{subsec:#1}}
\newcommand{\refeqshort}[1]{(\ref{eq:#1})}

\newcommand{\ignorethis}[1]{}
\newcommand{\norm}[1]{\lVert#1\rVert}

\usepackage{subcaption}
\usepackage{array}
\usepackage{multirow}
\usepackage[linesnumbered,ruled]{algorithm2e}
\usepackage{algpseudocode}
\usepackage{amsmath,amssymb,mathtools}
\usepackage{cuted}
\usepackage{amsthm}
\usepackage{afterpage}
\usepackage{xcolor}

\newif\ifsubmit

\submitfalse

\ifsubmit
\newcommand{\yl}[1]{\textcolor{red}{}}
\newcommand{\bo}[1]{\textcolor{blue}{}}
\newcommand{\chaowei}[1]{\textcolor{cyan}{}}
\newcommand{\dawei}[1]{\textcolor{yellow}{}}
\newcommand{\mingyan}[1]{\textcolor{purple}{}}
\newcommand{\jia}[1]{\textcolor{green}{}}
\newcommand{\wh}[1]{\textcolor{red}{}}
\newcommand{\kaizhao}[1]{\textcolor{cyan}{}}
\else
\newcommand{\yl}[1]{\textcolor{red}{Yulong: #1}}
\newcommand{\bo}[1]{\textcolor{blue}{Bo: #1}}
\newcommand{\chaowei}[1]{\textcolor{cyan}{Chaowei: #1}}
\newcommand{\dawei}[1]{\textcolor{orange}{Dawei: #1}}
\newcommand{\mingyan}[1]{\textcolor{purple}{Mingyan: #1}}
\newcommand{\jia}[1]{\textcolor{green}{Jia: #1}}
\newcommand{\wh}[1]{\textcolor{red}{[Warren: #1]}}
\newcommand{\kaizhao}[1]{\textcolor{cyan}{Kaizhao: #1}}

\fi
\newcommand{\StAdv}{\textit{LiDAR-Adv}\xspace}
\newcommand{\stAdv}{\textit{LiDAR-Adv}\xspace}

\makeatletter
\renewcommand{\paragraph}{%
  \@startsection{paragraph}{4}%
  {\z@}{0.5ex \@plus 1ex \@minus .2ex}{-1em}%
  {\normalfont\normalsize\bfseries}%
}
\makeatother
\begin{document}

\title{Adversarial Objects Against LiDAR-Based Autonomous Driving Systems}

\author{
Yulong Cao  \thanks{ The first three authors contributed equally.} $^{\ \, 1}$ \quad Chaowei  Xiao$^{*1}$ \quad Dawei Yang$^{*1}$  \quad Jing Fang $^{2}$ \quad Ruigang Yang $^{2}$ \\ \quad \textbf{Mingyan Liu$^{1}$ \quad Bo Li$^{3}$} \smallskip 
\\
$^{1}$University of Michigan, Ann Arbor\\
$^{2}$Baidu Research, Baidu Inc. \\
$^{3}$ University of Illinois at Urbana-Champaign\\
}

\maketitle
\begin{abstract}

Deep neural networks (DNNs) are found to be vulnerable against adversarial examples, which are carefully crafted inputs with a small magnitude of perturbation aiming to induce arbitrarily incorrect predictions. 
Recent studies show that adversarial examples can pose a threat to real-world security-critical applications: a ``physically adversarial \emph{Stop Sign}'' can be synthesized such that the autonomous driving cars will misrecognize it as others (e.g., a speed limit sign). %
However, these image-based adversarial examples cannot easily alter 3D scans such as widely equipped LiDAR or radar on autonomous vehicles. %
In this paper, we reveal the potential vulnerabilities of LiDAR-based autonomous driving detection systems, by proposing an optimization based approach \StAdv to
generate real-world adversarial objects that can evade the LiDAR-based detection systems under various conditions.
We first explore the vulnerabilities of LiDAR using an evolution-based blackbox attack algorithm, and then propose a strong attack strategy, using our gradient-based approach \stAdv.
We test the generated adversarial objects on the Baidu Apollo autonomous driving platform and show that such physical systems are indeed vulnerable to the proposed attacks. We 3D-print our adversarial objects and perform physical experiments with LiDAR equipped cars to illustrate the effectiveness of \StAdv. %
Please find more visualizations and physical experimental results on this website: \url{https://sites.google.com/view/lidar-adv}.

\end{abstract}

\section{Introduction}

Machine learning, especially deep neural networks (DNNs), have achieved great successes in various domains,~\citep{collobert2008unified,he2016deep,deng2013recent,silver2016mastering}.
Several safety-critical applications such as autonomous vehicles (AV) have also adopted machine learning models and achieved promising performance.
However, recent studies show that machine learning models are vulnerable to adversarial attacks~\cite{goodfellow2014explaining,carlini2016towards,xiao2018generating,xiao2018spatially,xiao2018characterizing,sun2018data}.
In these attacks, small perturbations are sufficient to cause various well-trained models to output ``adversarial" prediction.
In this paper we aim to explore similar vulnerabilities in today's autonomous driving systems.

Such adversarial attacks have been largely explored in the image domain.
In addition, to demonstrate such attacks pose a threat in the real world, some studies propose to generate physical stickers or printable textures that can confuse a classifier to recognize a stop sign~\cite{athalye2017synthesizing,evtimov2017robust}.
However, an autonomous driving system is not merely an image-based classifier.
For perception, most autonomous driving detection systems are equipped with LiDAR (Light Detection And Ranging) or RADAR (Radio Detection and Ranging) devices which are capable of directly probing the surrounding 3D environment with laser beams. This raises the doubt of whether texture perturbation in previous work will affect LiDAR-scanned point clouds.
In addition, the LiDAR-based detection system consists of multiple non-differentiable steps, rather than a single end-to-end network, which largely limits the use of gradient-based end-to-end attacks. 
These two key obstacles not only invalidate previous image-based approaches, but also raise several new challenges when we want to construct an adversarial object:
1) 
LiDAR-based detection system projects 3D shape to a point cloud using physical LiDAR equipment.
The point cloud is then fed into the machine learning detection system.
Therefore, how shape perturbation affects the scanned point cloud is not clear.
2) The preprocessing of the LiDAR point clouds is non-differentiable, preventing the naive use of gradient-based optimizers.
3) The perturbation space is limited due to multiple aspects. First, we need to ensure the perturbed object can be reconstructed in the real world. Second, a valid LiDAR scan of an object is a constrained subset of point cloud, making the perturbation space much smaller compared to perturbing the point cloud without any constraint~\cite{xiang2018generating}.

\begin{figure}[th]
\centering
 \includegraphics[width=0.85\linewidth]{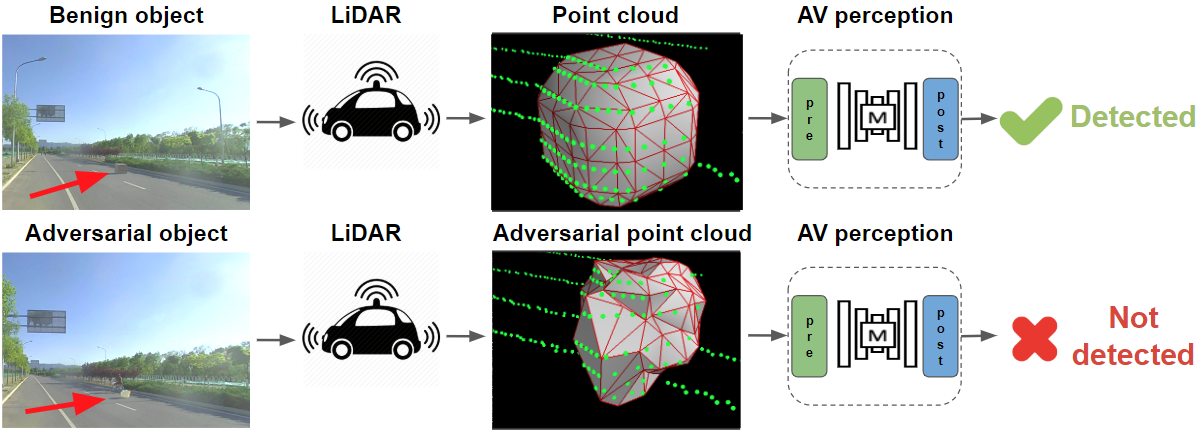}
 \caption{Overview of \stAdv. The first row shows that a normal box will be detected by the LiDAR-based detection system; while the generated adversarial object with similar size in row 2 cannot be detected.}
 \label{fig:overview}
\end{figure}

In this paper, we propose \StAdv to address the above issues and generate adversarial object against real-world LiDAR system as shown in Figure~\ref{fig:overview}. %
We first simulate a differentiable LiDAR renderer to bridge the perturbations from 3D objects
to LiDAR scans (or point cloud).
Then we formulate 3D feature aggregation with a differentiable proxy function.
Finally, we devise different losses to ensure the smoothness of the generated 3D adversarial objects.
To better demonstrate the flexibility of the proposed attack approach, we evaluate our attacking approach under two different attacking scenarios:
1) \emph{Hiding Object}: synthesizing an ``adversarial object'' that will not be detected by the detector.
2) \emph{Changing Label}: synthesizing an ``adversarial object'' that is recognized as a specified adversarial target by the detector.
We also compare \StAdv with the evolution algorithm in the blackbox setting.

To evaluate the real-world impact of \StAdv, we 3D print out the generated adversarial objects and test them on the Baidu Apollo autonomous driving platform, an industry-level system which is not only highly adopted for research purpose but also actively used in industries. We show that with 3D perception and a production-level multi-stage detector, we are able to mislead the autonomous driving system to achieve different adversarial targets.

To summarize, our contributions are as follows:
(1) We propose \stAdv, an end-to-end approach to generate physically plausible adversarial objects against LiDAR-based autonomous driving detection systems.
To the best of our knowledge, this is the first work to exploit adversarial objects for such systems.
(2)We experiment on Apollo, an industry-level autonomous driving platform, to illustrate the effectiveness and robustness of the attacks in practice. We also compare the objects generated by \StAdv with evolution algorithm to show that \StAdv can provide smoother objects.
(3) We conduct physical experiments by 3D-printing the optimized adversarial object and show that it can consistently mislead the LiDAR system equipped in a moving car.

\section{Related work}

\noindent\textbf{Image-space adversarial attacks}
Adversarial examples have been heavily explored in 2D image domains~\cite{goodfellow2014explaining,carlini2017towards,papernot2016limitations,moosavi2016deepfool,xiao2018generating}.
Various works~\cite{evtimov2017robust,kurakin2016adversarial,athalye2017synthesizing} start to study robust physical adversarial examples.
\citet{evtimov2017robust} has created printable 2D stickers to attach to a stop sign and cause a detector to predict wrong labels. 
Following this line, there are also works~\cite{yang2018realistic,liu2018adversarial} aiming to optimize the 3D shapes to show that even the surface geometry itself can produce adversarial behaviors.

In this work, we exploit the object surfaces to generate adversarial objects, and one fundamental challenge that differentiates our work from the previous ones is: the sensor in a LiDAR-based system directly probes the 3D environment as the input, bypassing surface textures of the adversarial objects. This means we may only rely on shape geometry to perform any attacks. On the other hand, compared to prior works that have shown successes on attacking single models, it is worth noting that the victim model which we experiment on (Apollo), is not merely an end-to-end deep learning model but an industry-level autonomous driving platform that consists of multiple non-differentiable parts.

\noindent\textbf{Adversarial point clouds}
~\citet{xiang2018generating} show a proof of concept, that models taking raw 3D point clouds as input~\cite{qi2017pointnet} can be vulnerable to adversarial point clouds. 
However, this approach is only evaluated with a single digital model.
It is not clear that the generated point clouds can form plausible 3D shape surfaces, or it can be reconstructed through LiDAR scans.
While in our approach, though the victim model takes point clouds as input similarly, these point clouds have to satisfy extra constraints such as: all points have to be the intersections of the laser beams and the object surfaces. We address this challenge by proposing a differentiable renderer which simulates the reconstructed laser beams projecting onto object surfaces. As we will show later, when the object moves, the point cloud changes in accordance with the laser hits, and how to enforce the robustness against such LiDAR scans is non-trivial.
\section{LiDAR-based Detection}
\begin{figure*}[t]
\centering
 \includegraphics[width=\linewidth]{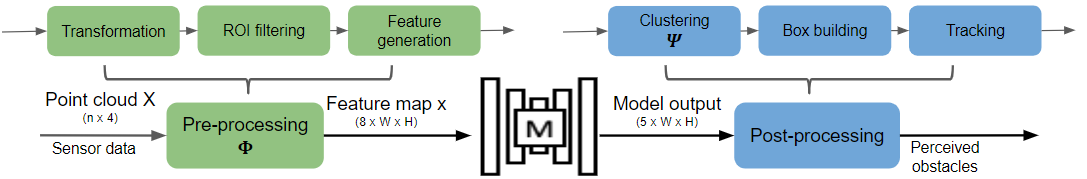}
 \caption{Overview of LiDAR-based detection on AV.}
 \label{fig:av-perception}
\end{figure*}

In this section, we provide the details of the LiDAR-based detection system that are directly related to our proposed adversarial attacks. Refer to the online repository\footnote{https://github.com/ApolloAuto/apollo/tree/r2.0.0/docs} for more details. %
\label{subsec:lidar-perception}
An overview of the system is shown in Fig.~\ref{fig:av-perception}.
First, a LiDAR sensor scans the 3D environment and obtains a raw point cloud of the scene.
Next, the point cloud goes through preprocessing, and is fed to a detection model.
Finally, post-processing is applied to the detection output to obtain the detection predictions.
\noindent\textbf{LiDAR.}
A LiDAR sensor scans the surrounding environment and generates a point cloud of $X \in \mathbb{R}^{n\times 4}$ with 3D coordinates ($u^{X},v^{X},w^{X}$) and intensity $int^{X}$. 
First, a sensor fires off an array of laser beams consecutive in horizontal and vertical directions. It then captures the light intensity reflecting back, and calculates the time that photons have traveled along each beam (Time of Flight). The distance and the coordinate of the surface points along the beam can be computed.
These points then form a raw point cloud of the object surfaces in the environment.
LiDAR sensors are supposedly robust to object surface textures, as the Time of Flight is not easily affected by texture change. Though it also detects the intensity of reflected lights, it is unclear how adversarial algorithms designed for natural lighting in image space can be adapted to invisible laser beams used as light sources.
Therefore, image-based adversarial attacks may have limited effects on such LiDAR-based detection system.
\paragraph{Preprocessing phase.}

\cut{
The previous raw point cloud $X$ goes through a preprocessing phase to form a feature map of $x\in \mathbb{R}^{H\times W\times 8}$ (see Sec.~\ref{sec:supp_background} for more details). %
Three components form this process: \emph{transformation}, \emph{Region of Interest (ROI) filtering} and \emph{feature generation}.

In the transformation process, the raw point cloud $X$ of size $n \times 4$ is transformed to an absolute coordinate system that a High Definition Map (HDMap) of the 3D environment is built upon.
Then, a ROI filter removes points from the point cloud $X$ which are not located above the road, to attain a ROI point cloud $X_{roi}$ based on HDMap information.}
The previous raw point cloud $X$ goes through a preprocessing phase to form a feature map of $x\in \mathbb{R}^{H\times W\times 8}$ (see Sec.~\ref{sec:supp_background}). The raw point cloud $X$ is first transformed and filtered based on a High Definition Map (HDMap) to attain a ROI point cloud $X_{roi}$.
This point cloud $X_{roi}$ is then sliced
into $H\times W$ vertical cells at 
$(\lfloor u^{X_{roi}}\rfloor, \lfloor v^{X_{roi}} \rfloor)$.
This ``hard'' assignment of points into cells will introduce piecewise zero gradients for \textbf{counting} and \textbf{max} \wrt the input. After slicing, in each cell, the information of the points are aggregated to generate a feature of size 8 for this cell, including heights, intensity, point counts \etc (detailed in Sec.~\ref{sec:supp_background})%
. %
This $H \times W \times 8$ feature map $x$ will then be fed into a machine learning model.

In this procedure, many operations (\eg max height, count) introduce zero gradients due to the ``hard'' assignment, so the end-to-end optimization-based attack algorithms are not directly applicable.

\paragraph{Machine learning model.}
Deep Neural Networks (DNNs) are used to process the $H\times W \times 8$ feature map, and then output the metrics for each one of the $H \times W$ cells. The metrics are listed in Sec.~\ref{sec:supp_background}.%

\paragraph{Post-processing phase.} The post-processing phase aggregates previous outputs from the machine learning model and recognizes the detected objects. The Post-processing can be roughly divided into 3 major sequential components: \emph{clustering}, \emph{box building} and \emph{tracking}. The clustering process composes obstacle candidates using both the model output metrics and ROI point cloud $X_{roi}$ generated from the preprocessing phase. In the clustering process, cells with higher \emph{objectness} confidence (greater than 0.5 by default) are used for constructing clusters by building a connected graph using \emph{center offset}. The obstacle candidates are produced by selecting the clusters with two constraints: (1) the average \emph{confidence} of cells in the cluster needs to be greater than 0.1 (2) the number of points in the ROI point cloud that are assigned to the cluster is greater than 3. The class probabilities of the obstacle candidate are calculated by summing up class probabilities of all cells in the cluster. The box builder then reconstructs the bounding boxes including the height, width, length of the obstacle candidates from the point cloud assigned to the candidate. Finally, the tracker integrates multiple frames of processed results to generate tracked objects as the output of the LiDAR-based detection, together with additional information such as object id, speed etc. 

Note that in this paper, we only consider a single frame for the adversarial attacks as a demonstration of feasibility. For the case of multiple frames, it can be treated as enhancing robustness against object motions, and such robustness against different locations is shown in later experiments (\shortrefsubsec{robust-adv}).
\section{Generating Adversarial Object Against LiDAR-based Detection}
In this section, we will formulate the problem first and describe the adversarial goals and challenges. We then describe our whitebox method \stAdv which aims to tackle the challenges and fulfill diverse adversarial goals. Finally, we propose an evolution-based attack method for blackbox settings.

\subsection{Methodology overview}
Given a 3D object $S$ in a scene, as stated in the background,
after the scene is scanned by a LiDAR sensor,
a point cloud $X$ is then generated based on $S$ so that $X = \mathrm{render}(S, \mathrm{background})
$
For preprocessing, this point cloud $X$ is sliced and aggregated to generate $x$, which is a $H\times W \times 8$ feature vector, and we call this aggregation process as $\Phi$:
$x=\Phi(X)$.
Then a machine learning model $M$ maps this 2D feature $x\in R^{H\times W \times 8}$ to $O=M(x)$, where $O \in R^{H\times W \times 7}$ (see Sec.~\ref{sec:supp_background} for concrete output meanings). $O$ is then post-processed by a clustering process $\Psi$ to generate the confidence $y_{conf}$ and label $y_{label}$ of detected obstacles so that $(y_{conf}, y_{label})=\Psi(O)$
An adversarial attacker aims to manipulate the object $S$ to achieve the adversarial goals. Here we define two types of adversarial goals: 1) \emph{Hiding object}: Hide an existing object $S$ by manipulating $S$; 2) \emph{Changing label}: Change the label $y$ of the detected object $S$ to a specified target $y'$. 

To achieve the above adversarial goals in LiDAR-based detection is non-trivial, and there are the following challenges:
1) \textbf{Multiple pre/post-processing stages.} Unlike the adversarial attacks on traditional image-space against machine learning tasks such as classification and object detection, the LiDAR-based detection here is not a single end-to-end learning model, It consists of the differentiable learning model $M$ and several non-differentiable parts including preprocessing and post-processing. Thus, the direct gradient based attacks are not directly applicable.
2) \textbf{Manipulation constraints.} %
Instead of directly manipulating the point cloud $X$ as in \cite{xiang2018generating}, we manipulate the 3D shape of $S$ given the limitation of LiDAR. The points in $X$ are the intersections of laser beams and object surfaces and cannot move freely, so the perturbations on each point may affect each other. Keeping the shape plausible and smooth adds additional constraints~\cite{yang2018realistic}.
3) \textbf{Limited Manipulation Space.}  
Consider the practical size of the object versus the size of the scene that is processed by LiDAR, the 3D manipulation space is rather small ($<2\%$ in our experiments), as shown in Fig.~\ref{fig:overview}.

Given the above challenges, we design an end-to-end attacking pipeline. In order to facilitate gradient-based algorithms, we implement an approximated differentiable renderer $R$ , which simulates the functionality of LiDAR, to intersect a set of predefined rays with the 3D object surface ($S$) consisting of vertices $V$ and faces $W$. The points at the intersections form the raw point cloud $X$. After preprocessing, the point cloud is then fed to a preprocessing function $\Phi$ to generate the feature map $x = \Phi(X)$. The feature map $x$ is then taken as input for a machine learning model $M$ to obtain the output metrics $O = M(x)$.

The whole progress can be symbolized as $F(S) = M(\Phi(R(S)))$. Note that by differentiating the renderer $R$, the whole process $F(\ast) = M(\Phi(R(\ast)))$ is differentiable w.r.t.\@ $S$. In this way, we can manipulate $S$ to generate adversarial $S_{\adv}$ via our designed objective function operating on the final output $F(S)$. 

\subsection{Approximate differentiable renderer}
\label{subsec:diff_render}
\paragraph{LiDAR simulation}
The renderer $R$ simulates the physics of a LiDAR sensor that probes the objects in the scene by casting laser beams. The renderer first takes a mesh $S$ as input, and compute the intersections of a set of predefined ray directions to the meshes in the scene to generate point cloud $X$. After depth testing, the distance along each beam is then captured, representing the surface point of a mesh that it first encounters, as if a LiDAR system receives a reflection from an object surface. Knowing ray directions of the beams, the exact positions of the intersection points can be inferred from the distance, in the form of point clouds $X$.

\paragraph{Real background from a road scene} 
We render our synthetic object onto a realistically captured point cloud. First, we obtain the 3D scan of a road scene, using the LiDAR sensor Riegl VMX-1HA mounted on a car. Then, we obtain the laser beam directions by computing the normalized vectors from the origin (LiDAR) pointing to the scanned points.
This fixed set of ray directions are then used for rendering our synthetic objects throughout the paper. Note that we can also manually set ray directions given sensor specifications, but it will be less real, because it may not model the noises and fluctuations that occur in a real LiDAR sensor.

\paragraph{Hybrid rendering of synthetic objects onto a realistic background}
Given the ray directions reconstructed from the background point cloud, a subset will intersect with the object, forming the point cloud for the object of interest.
The corresponding background points are then removed since these background points are occluded by the foreground object.
In this way, we obtain a semi-real synthetic point cloud scene: the background points come from the captured real data; the foreground points are physically accurate simulated based on the collected real data.

\subsection{Differentiable proxy function for feature aggregation}
As in Section~\ref{subsec:lidar-perception}, in the preprocessing of Apollo, it aggregates the point cloud into hardcoded 2D features, including \textbf{count}, \textbf{max height}, \textbf{mean height}, \textbf{intensity} and \textbf{non-empty}. These operations are non-differentiable. In order to apply end-to-end optimizers to for our synthetic object $S$, we need to flow the gradient through the feature aggregation step, with the help of our proxy functions.

Given a point cloud $X$ with coordinate $(u^{X}, v^{X}, w^{X})$, and we hope to count the number of points falling into the cells of a 3D grid $G\in\mathbb{R}^{H\times W \times P}$. For a point $X_i$ with location$(u^{X_i}, v^{X_i}, w^{X_i})$,
we increase the count of 8 cells: the centers of these 8 cells form a cube, and the point $X_i$ is inside this cube.
Specifically, we increase the count of 8 cells using trilinear weights:
\begin{small}
\begin{equation} \label{eq1}
\begin{split}
   & G(u_i, v_i, w_i) =  \sum_{p}(1 - d(u_p, u^{X_i})) \cdot(1 - d(v_p , w^{X_i}))\cdot (1 - d(w_p, w^{X_i})), \\
\end{split}
\end{equation}
\end{small}
where  $p \in \mathcal{N}(u^{X_i}, v^{X_i}, w^{X_i})$ are the indices of the 8-pixel neighbors at location $(u^{X_i}, v^{X_i}, w^{X_i})$ and $d(\cdot, \cdot)$ represents the $L_1$ distance.
The $\textbf{count}$ feature $x_{\mathbf{count}}$ is the value $G_p=G(u_i,v_i,w_i)$ computed for each grid $i$.
Note that this feature is no longer an integer and can have non-zero gradients \wrt the point coordinates.

We then use this ``soft count'' feature to further compute ``mean height'' and ``max height'' features.
For simplicity, we first define a constant height matrix $T\in\mathcal{R}^{H, W, P}$, where $T(.,.,p) = p, p\in \{1...P\}$. This matrix stores the height of each cell. Next, we can formulate the \textbf{mean height} $x_{\mathbf{mean-height}}$ and \textbf{max height} $x_{\mathbf{max-height}}$ using soft count $G$:
\begin{small}
\begin{equation}
   x_{\mathbf{mean\xspace-height}} = \cdot  \frac{  \sum_{p\in P}{G_p \circ T_p}}{ \sum_{p\in P}{G_p} + \epsilon}\qquad\mathrm{and}\qquad
x_{\mathbf{max\xspace-height}} =\max_{p}{\sign{(G(.,.,p))}  \circ T(.,.,p)}
\end{equation}
\end{small}
where $\epsilon=1e^{-7}$ to prevent zero denominators.
Note that the $\sign$ function is non-differentiable, so we approximate the gradient using $\sign(G) = G$ during back propagation. The feature \textbf{intensity} has the similar formulation of \textbf{height} so we omit them here. The feature $\textbf{non-empty}$ is formulated as $x_{\mathbf{non-empty}} = \sign(G)$.

We denote the above trilinear approximator as $\Phi'$, in constrast to the original non-differentiable preprocessing step $\Phi$. A visualization of output of our $\Phi'(X)_{\mathbf{count}}$ compared to the original $\Phi(X)_{\mathbf{count}}$ is shown in Sec.~\ref{sec:supp_grad_proxy_func}. Since our approximation introduces differences in counting, $\Phi'(X)$ is not strictly equal to $\Phi(X)$, resulting in different $\mathrm{obj}$ values of the final model prediction. We observe that this difference will raise new challenges to transfer the adversarial object generated based on $\Phi'$ to $\Phi$. To solve this problem, we reduce the difference between $\Phi'$ and $\Phi$, by replacing the L1 distance $d$ in Eq.~\ref{eq1} with $d(u_1, u_2) = 0.5 + 0.5\cdot\tanh(\mu\cdot(u_1 - u_2-1))$ where $\mu = 20$. We name this approximation ``tanh approximator'' and denote it $\Phi''$. We observe that the input difference between $\Phi''$ and the original $\Phi$ is largely reduced compared to $\Phi'$, allowing for smaller errors of the model predictions and better transferability.
To extend our approximator and further reduce the gap betweer $\Phi''$ and $\Phi$, we interpolate the distance: $d(u_1,u_2)=\alpha \cdot (0.5 + 0.5\cdot\tanh(5\mu\cdot(u_1 - u_2-1)))  + (1-\alpha)\cdot (u_1 - \lfloor u_2 \rfloor )$, where $\alpha$ is a hyper-parameter balancing the accuracy of approximation and the availability of gradients.
\subsection{Objective functions}
Our objective is to generate a synthetic adversarial object $S^{\adv}$ from an original object $S$ by perturbing its vertices, such that the LiDAR-based detection model will make incorrect predictions.
We first optimize $S^{\adv}$ against the semi-real simulator detection model $M$.
\begin{equation}\label{eq:lambda_select}
    \gL(S^{\adv}) =  \gL_{\adv}(S^{\adv}, M) + \lambda \gL_{\prep}(S^{\adv}; S)
\end{equation}
The objective function $\gL$ consists of two losses. $\gL_{\adv}$ is the adversarial loss to achieve the target goals while the $\gL_{\prep}$ is the distance loss to keep the properties of the ``realistic" adversarial 3D object $S^{\adv}$. We optimize the objective function by manipulating the vertices.
The distance loss is defined as follows:
\begin{small}
\begin{equation}
    \gL_{\prep}  =  \sum_{ \vv_i \in V} \sum_{q \in{\mathcal{N}(\vv_i)}} \norm{\Delta\vv_i - \Delta\vv_q}^2_2 + \beta \sum_{\vv_i \in V} \norm{\Delta\vv_i}^2_2,
\end{equation} 
\end{small}
where $\Delta\vv_i=\vv_i^\adv - \vv_i$
represents the displacement between the adversarial vertex $\vv_i^\adv$  and pristine vertex $\vv_i$. $\beta$ is the hyperparameter balancing these two losses. The first losses~\cite{yang2018realistic} is a Laplacian loss preserving the smoothness of the perterbation applied on the adversarial object $S^{adv}$.
The second part is the $L2$ distance loss to limit the magnitude of perturbation. 
\cut{
For example: the obstacle of the car by adding the spoofed sphere in Fig.~\ref{fig:overview} without $L_2$ loss, the position of the spoofed object may shift to vacant positions without the car at its' bottom and violates physics, since shifting has no effect on the Laplacian loss.
Based on different adversarial targets, we designed the objective function as follows respectively.
}
\paragraph{Objective: hide the inserted adversarial object}
As introduced in the background section, the existence of the object highly depends on the ``positiveness'' metric. $H(\ast, M,  S)$ denotes a function extracting $\ast$ metric from the model $M$ given an object $S$.  $\mathcal{A}$ is the mask of the target object's bounding box. Our adversarial loss is represented as follows:
\begin{equation}
    \gL_{\adv} = H(\mathrm{pos}, M, S) * \mathcal{A}
\end{equation} 
\paragraph{Objective: changing label}
In order to change the predicted label of the object, it needs to increase the logits of the target label and decrease the logits of the ground-truth label. Moreover, it also needs to preserve the high positiveness. Based on this, our adversarial loss is written as
\begin{equation}
\gL_{\adv} = (-H(\mathrm{c}_{y'}, M, S) + H(\mathrm(c)_y, M, S ) \cdot \mathcal{A} *  H(\mathrm{pos}, M, S)
\end{equation}
In order to ensure that adversarial behaviors still exist when the settings are slightly different, we create robust adversarial objects that can perform successful attacks within a range of settings, such as different object orientations, different positions to the LiDAR sensor \etc. To achieve such goal, we sample a set of physical transformations to optimize the loss expectation. In reality, we create a victim set $D$ by rendering the object $S$ at different positions and orientations. Instead of optimizing an adversarial object $S$ by attacking single position and orientation, we generate an universal adversarial object $S$ to attack all positions and orientations in the victim set $D$.
\subsection{Blackbox Attack}
In reality, it is possible that the attackers do not have complete access to the internal model parameters, \ie the model is a black box. Therefore, in this subsection, we also develop an evolution-based approach to perform blackbox attack.

In evolution, a set of individuals represent the solutions in the search space, and the fitness score defines how good the individuals are.
In our case, the individuals are mesh vertices of our adversarial object, and the fitness score is $-\gL(S^{\adv})$. We initialize $m$ mesh vertices using the benign object $S$. For each iteration, new population of $n$ mesh vertices are generated by adding random perturbations, drawn from a Gaussian distribution $\mathcal{N}(0, \sigma)$, to each mesh vertices in the old population. $m$ mesh vertices with top fitness scores will remain for the next iteration, while the others will be replaced. We iterate the process until we find a valid solution or reach a maximum number of steps.

\section{Experiments}
In this section, we first expose the vulnerability of the LiDAR-based detection system via the evolution-based blackbox algorithm by achieving the goal of ``hiding object'', because missing obstacles can cause accidents in real life. We then show the qualitative results and quantitative results of \StAdv under whitebox settings. In addition, we also show that \StAdv can achieve other adversarial goals such as  ``changing label''.
Moreover, the point clouds are continuously captured in real life, so attacks in a single static frame may not have much effect in real-world cases.
Therefore, in our experiments, we generate a universal robust adversarial object against a victim dataset which consists of different orientations and positions. We 3D-print such universal adversarial object and conduct the real-world drive-by experiments, to show that they indeed can pose a threat on road.

\subsection{Experimental setup}
We conduct the evaluation on the perception module of Baidu Apollo Autonomous Driving platform (V2.0). 
We initialize the adversarial object as a resampled 3D cube-shaped CAD model using MeshLab~\cite{cignoni2008meshlab}.
For rendering, we implement a fully differential LiDAR simulator with predefined laser beam ray directions extracted from a real scene captured by the Velodyne HDL-64E sensor, as stated in~\shortrefsubsec{diff_render}. It has around 2000 angles in the azimuth angle and around 60 angles in the elevation angle. %
We use Adam optimizers~\cite{kingma2014adam}, and choose $\lambda$ as $0.003$ in Eq.~\ref{eq:lambda_select} using binary search. For the evolution-based blackbox algorihtm , we choose $\sigma=0.1$, $n=500$ and $m=5$.

\subsection{Vulnerability analysis}
Here, we first show the existence of the vulnerability using our evolution-based blackbox attacks, with the goal of ``hiding object''. We generate adversarial objects in different size (50cm and 75cm in edge length). For each object, we select 45 different position and orientation pairs for evaluation, and the results are shown in Table~\ref{tbl:single-quant}. The results indicate that the LiDAR-based detection system is vulnerable. The visualization of the adversarial object is shown in Figure~\ref{fig:size}(a) and (c).
\subsection{\stAdv with different adversarial goals}
\begin{figure}[h]
    \centering
    \begin{minipage}{.15\textwidth}
     \begin{subfigure}{\textwidth}
     \centering
     \includegraphics[width=\textwidth]{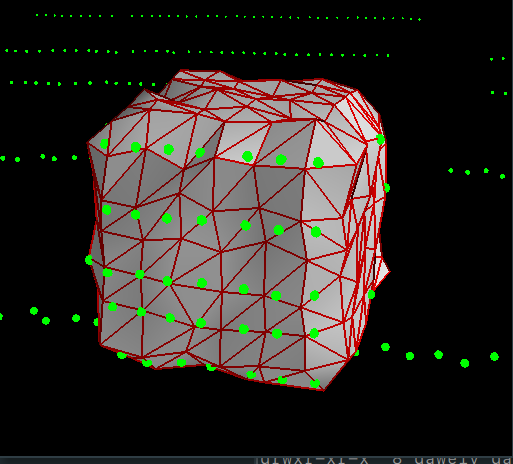}
     \caption{50cm: Evo.}
     \label{fig:50_de}
     \end{subfigure}
    \end{minipage}~
    \begin{minipage}{.15\textwidth}
     \begin{subfigure}{\textwidth}
     \centering
     \includegraphics[width=\textwidth]{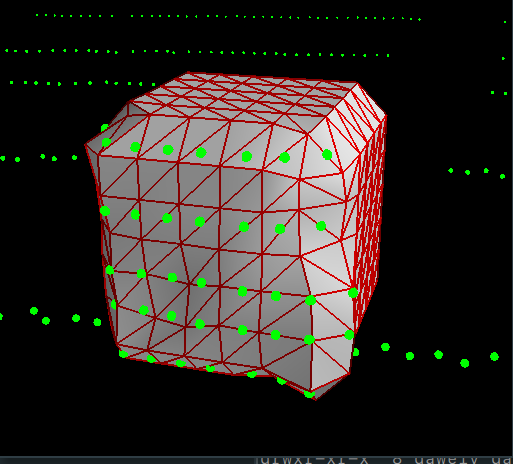}
     \caption{ \StAdv}
     \label{fig:50}
     \end{subfigure}
    \end{minipage}~
    \begin{minipage}{.15\textwidth}
     \begin{subfigure}{\textwidth}
     \centering
     \includegraphics[width=\textwidth]{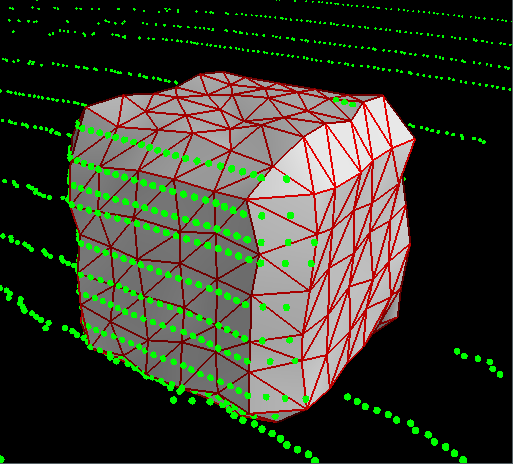}
     \caption{75cm: Evo.}
     \label{fig:75_de}
     \end{subfigure}
    \end{minipage}~
   \begin{minipage}{.15\textwidth}
     \begin{subfigure}{\textwidth}
     \centering
     \includegraphics[width=\textwidth]{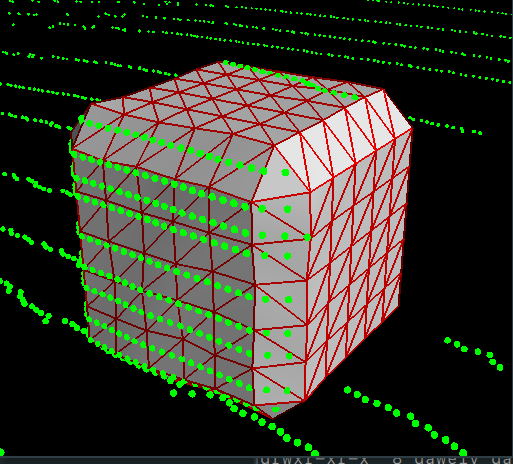}
     \caption{\StAdv}
     \label{fig:75}
     \end{subfigure}
    \end{minipage}~
    \caption{Adversarial meshes of different sizes can fool the detectors even with more LiDAR hits. We generate the object with  \StAdv and evolution-based method (Evo.).}
    \label{fig:size}
\end{figure}
After showing the vulnerability of the LiDAR-based detection system, here we focus on whitebox settings to explore what a powerful adversary can do, since ``the design of a system should not require secrecy''~\citep{shannon1949communication}.
Therefore, we evaluate the effectiveness of our whitebox attack \StAdv with the goal of ``hidding object''. We also evaluate the feasibility of \StAdv to achieve another goal of ``changing label''.
\paragraph{Hiding object} 
\begin{wraptable}{r}{0.5\textwidth}
\caption{Attack success rate of \StAdv and evolution based method under different settings.}
\label{tbl:single-quant}
\begin{minipage}{.5\textwidth}
\resizebox{.95\columnwidth}{!}{%
\begin{tabular}{l| l | l} 
\toprule
\multirow{2}{*}{Attacks} &  \multicolumn{2}{c}{Object size}\\
 & 50cm & 75cm  \\
\midrule
\StAdv & 32/45 (71\%) & 23/45 (51\%) \\
Evolution-based & 28/45 (62\%) & 16/45 (36\%)\\
\bottomrule
\end{tabular}
}
\end{minipage}~
\end{wraptable}
We follow the same settings as in the above sections, and Table~\ref{tbl:single-quant} shows the results. We find that \StAdv can achieve 71\% attack success rate with size 50cm. The attack success rate is consistently higher than the evolution-based blackbox attacks. Figure~\ref{fig:size} (b) and (c) show the visualizations of the adversarial objects. We visually observe that the adversarial objects generated by \StAdv are smoother than that of evolution.

\begin{figure*}[h!]
\centering
\begin{minipage}{0.49\linewidth}
 \begin{subfigure}{.30\linewidth}
 \includegraphics[width=\linewidth]{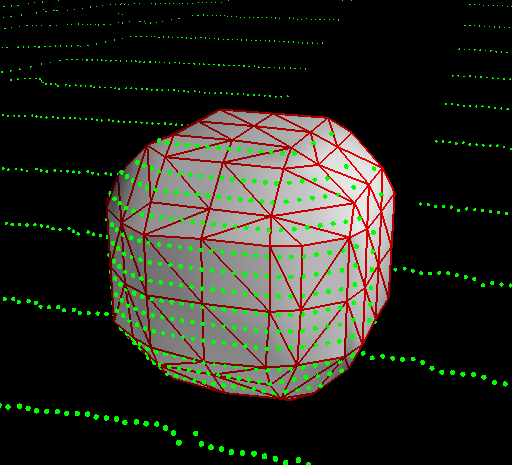}
 \caption{3D mesh \\ \;\; (Benign)}
 \end{subfigure}
  \begin{subfigure}{.68\linewidth}
 \includegraphics[width=\linewidth]{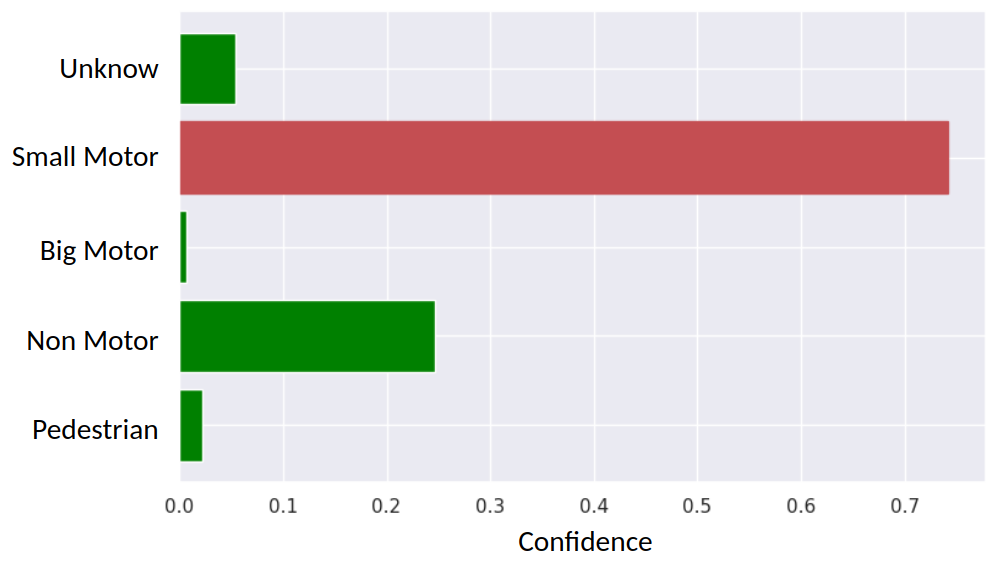}
 \caption{Predictions (Benign)}
 \end{subfigure}
\end{minipage}
\begin{minipage}{0.49\linewidth}
  \begin{subfigure}{.30\linewidth}
 \includegraphics[width=\linewidth]{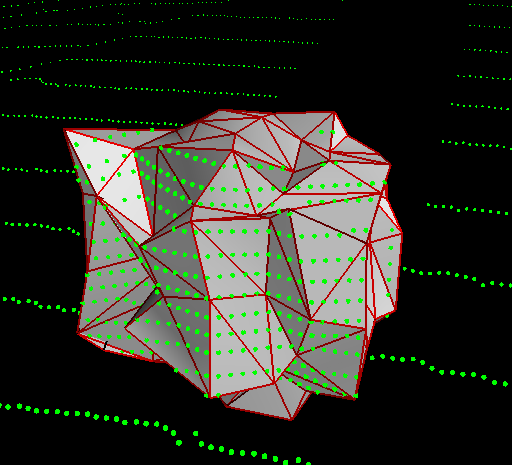}
 \caption{3D Mesh \\ \;\; (Adv.)}
 \end{subfigure}
  \begin{subfigure}{.68\linewidth}
\includegraphics[width=\linewidth]{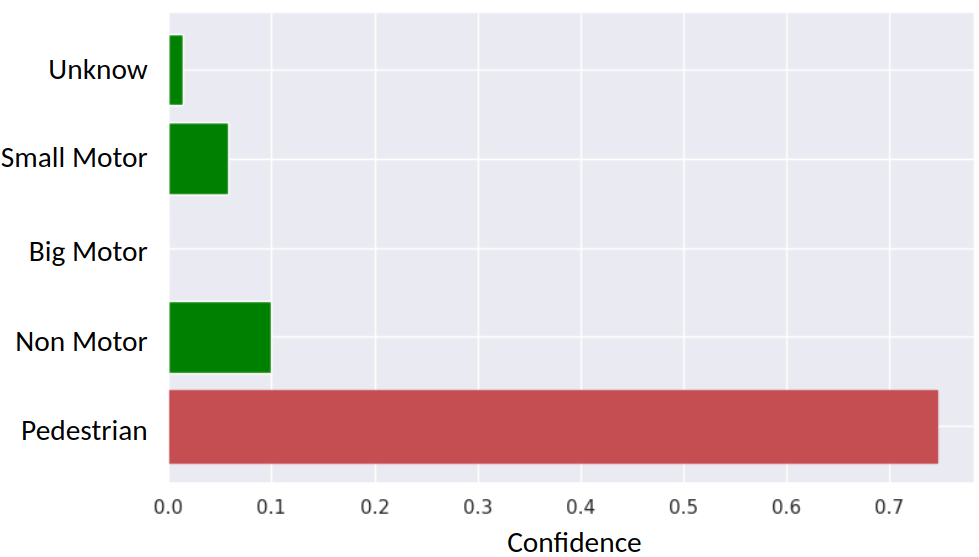}
 \caption{Predictions (Adv.)}
 \end{subfigure}
\end{minipage}
\caption{The adversarial mesh generated by \StAdv is mis-detected as a ``Pedestrian''.}
\label{fig:change-label2}
\end{figure*}

\paragraph{Changing label} %
The result shown in Figure~\ref{fig:change-label2} indicates that we can successfully change the label of the object. We also experiment with different initial shapes and target labels. More details can be found in Sec.~\ref{sec:supp_additional_result}.

\subsection{\stAdv on generating robust physical adversarial objects}
\label{subsec:robust-adv}
To ensure the generated \StAdv preserves adversarial behaviors under various physical conditions, we optimize the object by sampling a set of physical transformations such as possible positions and orientations.
We show that the generated robust adversarial object is able to achieve the attack goal of hiding object with a high success rate in Table~\ref{tbl:all}. An interesting phenomenon is that some attack performance under the unseen settings is even better than that within the controlled environment. This implies that our adversarial objects are robust enough to generalize to unseen settings.
\begin{table}
\caption{Attack success rates of \StAdv at different positions and orientations under both controlled and unseen settings. 
}
\label{tbl:all}
\centering
\begin{tabular}{l l| l l | l l   } 
\toprule
\multicolumn{2}{c|}{ Controlled Setting } &  \multicolumn{4}{c}{Unseen Setting}\\
\multirow{2}{*}{\shortstack{Distance (cm) \& Orientation (\degree)}}& \multirow{2}{*}{Attack}& \multicolumn{2}{c|}{Distance (cm)} & \multicolumn{2}{c}{Orientation (\degree)}  \\
& & 0-50 & 50-100 & 0-5 & 0-10\\
\midrule
$\{0,\pm50\}\times\{0, \pm2.5, \pm5\}$ & 41/45  & 96/100 & 91/100 & 10/10 & 9/10 \\
$\{0,\pm50\}\times\{0, \pm2.5, \pm5\}$ & 43/45  & 96/100 & 90/100 & 8/10 & 10/10 \\
\bottomrule
\end{tabular}
\end{table}

\begin{figure}[h]
    \centering
    \begin{minipage}{.48\textwidth}
     \begin{subfigure}{\textwidth}
     \centering
     \includegraphics[width=\textwidth]{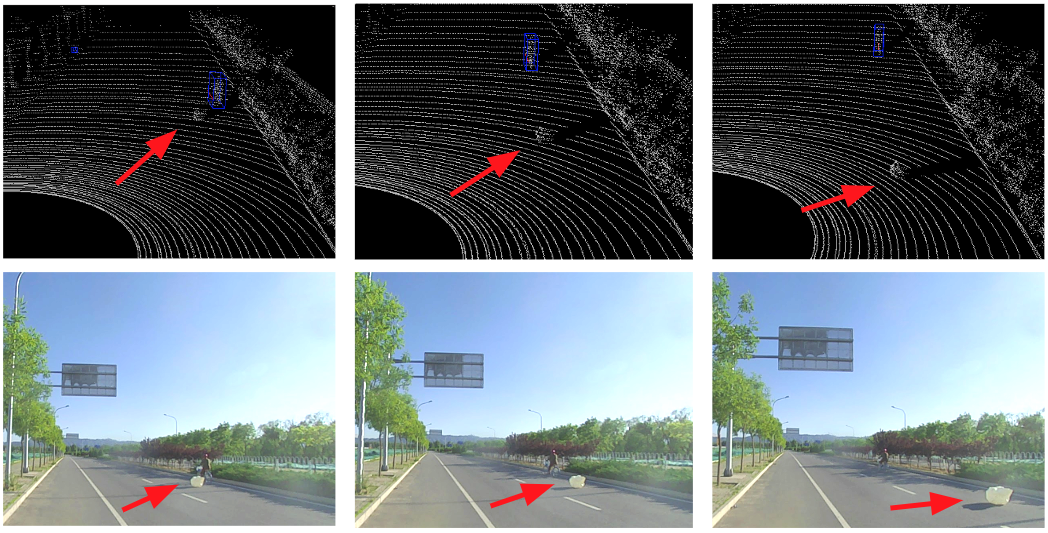}
     \caption{Adversarial object}
     \label{fig:adv_phy}
     \end{subfigure}
    \end{minipage}~\vrule~\begin{minipage}{.48\textwidth}
     \begin{subfigure}{\textwidth}
     \centering
     \includegraphics[width=\textwidth]{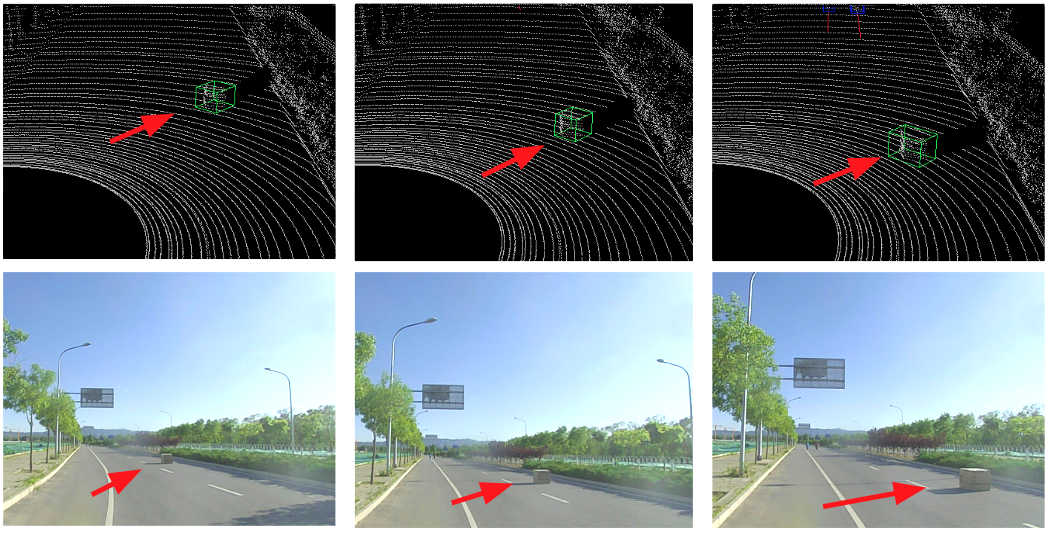}
     \caption{Benign object}
     \label{fig:box_phy}
     \end{subfigure}
    \end{minipage}~
    \caption{Results of physical attack. Our 3D-printed robust adversarial object by \StAdv is not detected by the LiDAR-based detection system in a moving car. Row 1 shows the point cloud data collected by LiDAR sensor, and Row 2 presents the corresponding images captured by a dash camera.}
    \label{fig:phy}
\end{figure}

Furthermore, we evaluate the generated robust adversarial object in the physical world by 3D printing the generated object. We collect the point cloud data using a Velodyne HDL-64E sensor with a real car driving by and evaluate the collected traces on the LiDAR perceptual module of Baidu Apollo. As shown in Figure~\ref{fig:adv_phy}, we find that the adversarial object is not detected around the target position among all 36 different frames. To compare, the box object (in Figure~\ref{fig:box_phy}) is detected in 12 frames among all 18 frames. The number of total frames is different due to the different vehicle speed. More details can be found in Sec.~\ref{sec:supp_additional_result}. 

\section{Conclusion}
We show that LiDAR-based detection systems for autonomous driving are vulnerable against adversarial attacks.
By integrating our proxy differentiable approximator, we are able to generate robust physical adversarial objects.
We show that the adversarial objects are able to attack the Baidu Apollo system at different positions with various orientations. We also show \StAdv can generate much smoother object than evolution based attack algorithm.
Our findings raise great concerns about the security of LiDAR systems in AV, and we hope this work will shed light on potential defense methods.

\newpage

{\small
\setlength{\bibsep}{0pt}
\bibliographystyle{abbrvnat}
\bibliographystyle{ieee}
\bibliography{main}
}
\appendix

\renewcommand\thefigure{\Alph{figure}}
\renewcommand\thetable{\Alph{table}}
\renewcommand\thesection{\Alph{section}}

\newpage
\section{Differential Renderer}\label{sec:supp_diff_render}
\paragraph{LiDAR Simulation}
The renderer simulates the physics of a LiDAR sensor that probes the objects in the scene by casting laser $N_{\mathrm{ray}}$ rays: $R = \{\mathbf{r}_i\in\mathbb{R}^3, \|\mathbf{r}_i\|=1,i=1,2,\cdots,N_\mathrm{ray}\}$, with $\mathbf{r}_i$ representing the direction of the $i$-th ray.
Given a shape $S$ with the surface $\partial S$ as input, the renderer computes the intersections of rays $R$ to the mesh faces in the scene. For each ray $\mathbf{r}_i$, the intersection coordinate $P_i$ are computed through
depth testing (assuming the center of the rays is at origin, \ie the reference frame of LiDAR):
\begin{equation}\label{eq:render}
\begin{split}
    \mathbf{p}_i = \argmin_{\mathbf{p}} \{\|\mathbf{p}\|\ |\  \exists t>0, \mathbf{p} = t\cdot\mathbf{r}_i, \mathbf{p} \in \partial S\}, \\
    \ i=1,2,\cdots,N_\mathrm{ray}
\end{split}
\end{equation}

\paragraph{Object insertion}
Notice that we have a predefined set of rays $R$.
To obtain these rays, one can refer to the specifications of a LiDAR device. In our paper, we directly compute the directions from the captured background point cloud $P'$, so that the rays are close to real world cases:
\begin{equation}
    \mathbf{r_i} = \frac{\mathbf{p}'_i}{\|\mathbf{p}'_i\|}
\end{equation}
With this, Eq.~\refeqshort{render} becomes:
\begin{equation}\label{eq:render2}
\begin{split}
    \mathbf{p}_i = \argmin_{\mathbf{p}}\{\|\mathbf{p}\|\ |\ \mathbf{p} = \mathbf{p}'_i \vee \mathbf{p} = t\cdot\mathbf{r}_i, t>0, \mathbf{p} \in \partial S\}, \\
    \ i=1,2,\cdots,N_\mathrm{ray}
\end{split}
\end{equation}
This means when rays intersect with an object, the corresponding background points blocked by the above-ground parts of the object are removed during depth testing; if the object is below the ground, the intersections leave those corresponding background points intact also due to depth testing.
In this way, we obtain a semi-real synthetic point cloud scene: the background points come from the captured real data; the foreground points are physically accurate simulations based on the captured real data.

\section{Background}\label{sec:supp_background}
\subsection{LiDAR perception system}
Detailed machine learning model input features and machine learning model output metrics are shown in Table~\ref{table:cnn_inputoi} and Table~\ref{table:cnn_output}.
\begin{table}[h]\footnotesize
  \caption{Machine learning model input features extracted in the preprocessing phase.}
  \label{table:cnn_inputoi}
\centering
\begin{tabular}{ l | l }
\toprule
\textbf{Feature} & \textbf{Description} \\
\midrule
\textbf{Max height} & Maximum height of points in the cell. \\
\textbf{Max intensity} & Intensity of the highest point in the cell. \\
\textbf{Mean height} & Mean height of points in the cell. \\
\textbf{Mean intensity} & Mean intensity of points in the cell. \\
\textbf{Count} & Number of points in the cell. \\
\textbf{Direction} & Angle of the cell’s center with respect to the origin. \\
\textbf{Distance} & Distance between the cell’s center and the origin. \\
\textbf{Non-empty} & Binary value indicating whether the cell is empty or occupied. \\
\bottomrule
\end{tabular}
\\
\end{table}

\begin{table}[h]\footnotesize
  \caption{Output metrics of the segmentation model.}
  \label{table:cnn_output}
  \centering
\begin{tabular}{ l | l }
\toprule
\textbf{Metric} & \textbf{Description} \\
\midrule
\textbf{Center offset ($\mathrm{off}$)} & Offset to predicted center of the cluster the cell belongs to. \\
\textbf{Objectness ($\mathrm{obj}$)} & The probability of a cell belonging to an obstacle. \\
\textbf{Positiveness ($\mathrm{pos}$)} & The confidence score of the detection. \\
\textbf{Object height ($\mathrm{hei}$)} & The predicted object height. \\
\textbf{$i\-$th Class Probability  ($\mathrm{cls_i}$)} & The probability of the cell being from class $i$ (vehicle, pedestrian, etc.). \\
\bottomrule
\end{tabular}
\\
\end{table}

\section{Generating Adversarial Object Against LiDAR Perception}\label{sec:supp_grad_proxy_func}
\subsection{Gradient of proxy functions}
\begin{figure*}[t]
    \centering
    \begin{minipage}{0.24\linewidth}
     \begin{subfigure}{\linewidth}
     \centering
     \includegraphics[width=\linewidth]{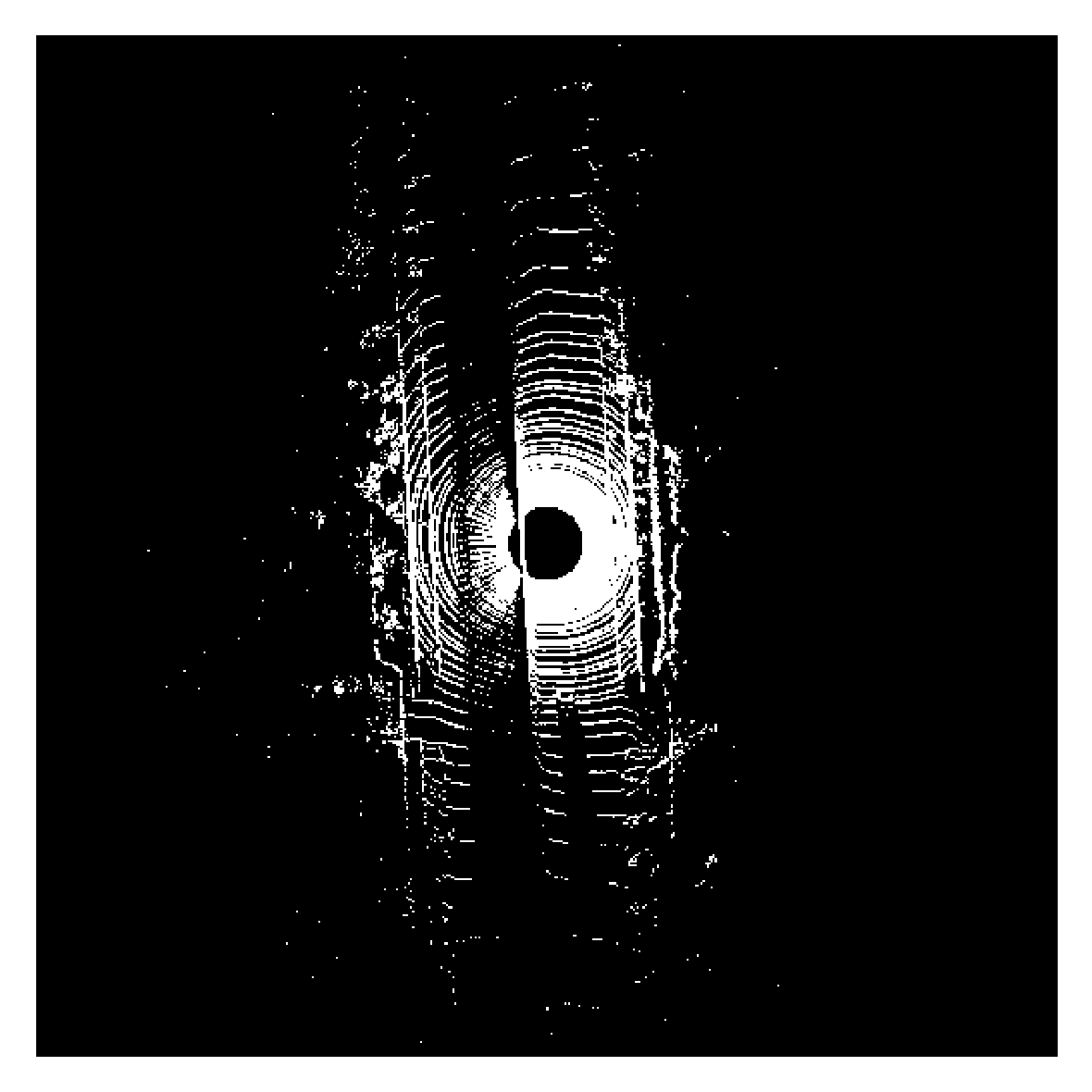}
     \caption{$\Phi'(X)_{\mathbf{count}}$ }
     \label{fig:img_benign}
     \end{subfigure}
    \end{minipage}
    \begin{minipage}{0.24\linewidth}
     \begin{subfigure}{\linewidth}
     \centering
     \includegraphics[width=\linewidth]{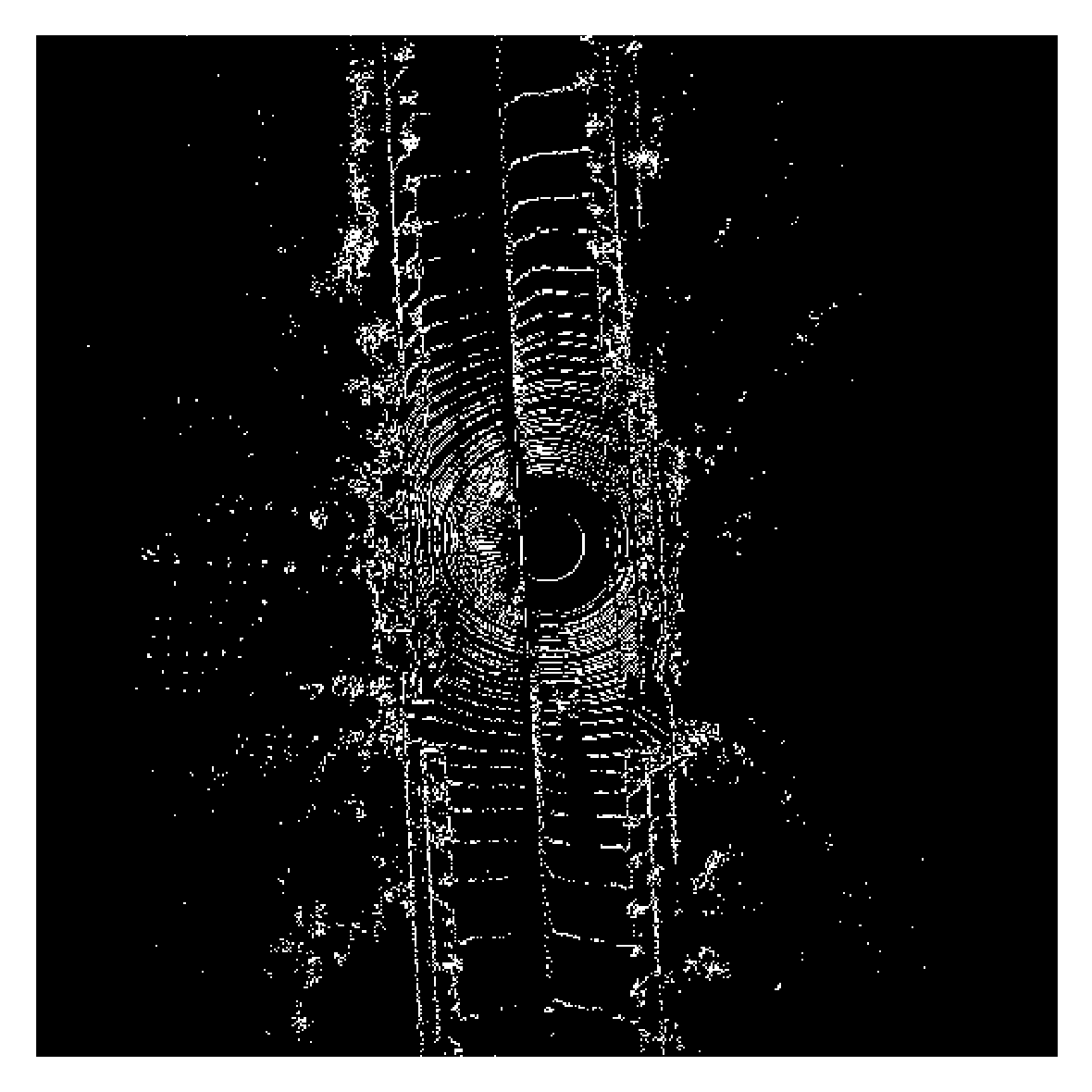}
     \caption{\scalebox{0.7}{$\Phi'(X)_{\mathbf{count}}$ - $\Phi(X)_{\mathbf{count}}$}}
     \label{fig:img_benign}
     \end{subfigure}
    \end{minipage}
    \begin{minipage}{0.24\linewidth}
     \begin{subfigure}{\linewidth}
     \centering
     \includegraphics[width=\linewidth]{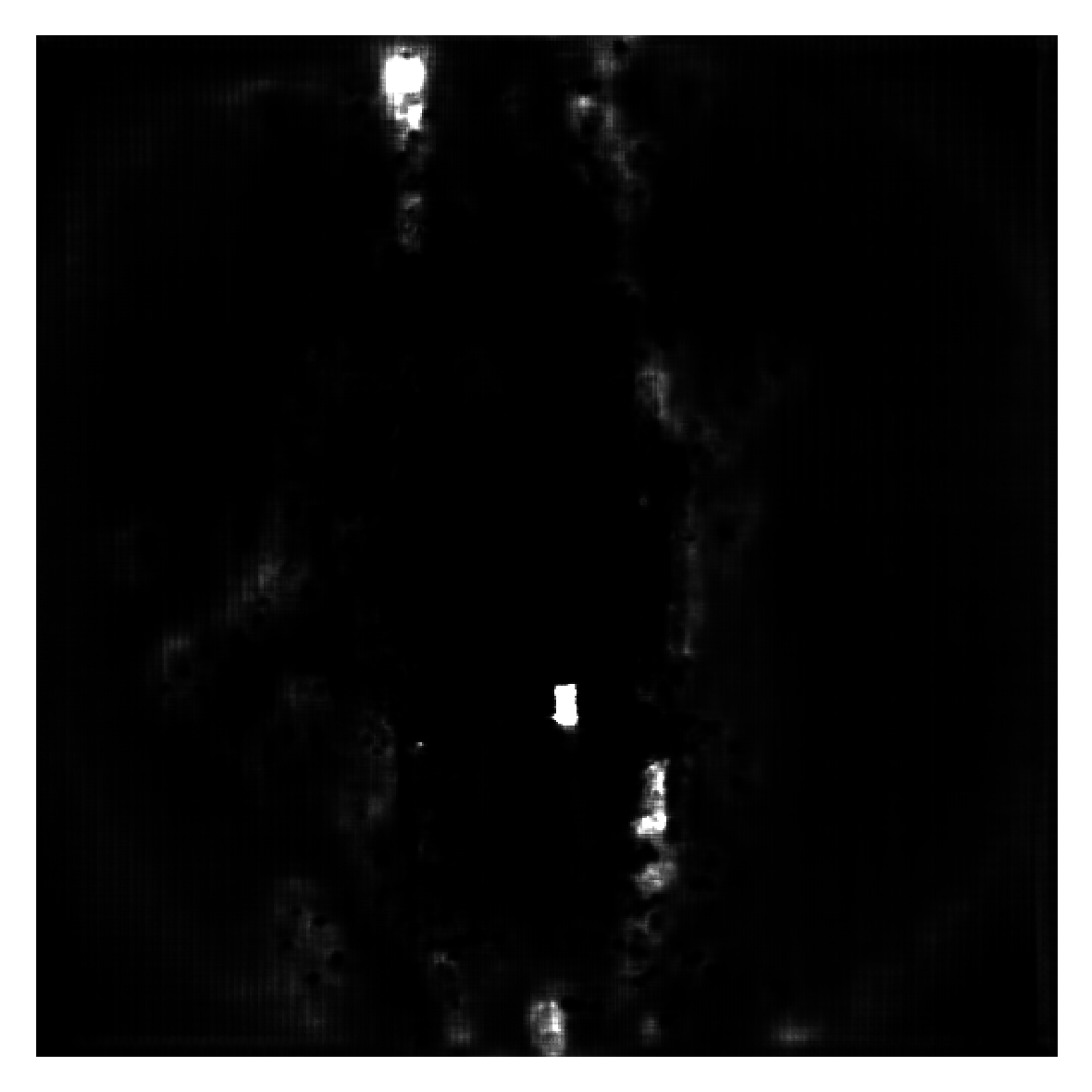}
     \caption{$M(\Phi'(X))_{\mathbf{obj}}$}
     \label{fig:img_benign}
     \end{subfigure}
    \end{minipage}
    \begin{minipage}{0.24\linewidth}
     \begin{subfigure}{\linewidth}
     \centering
     \includegraphics[width=\linewidth]{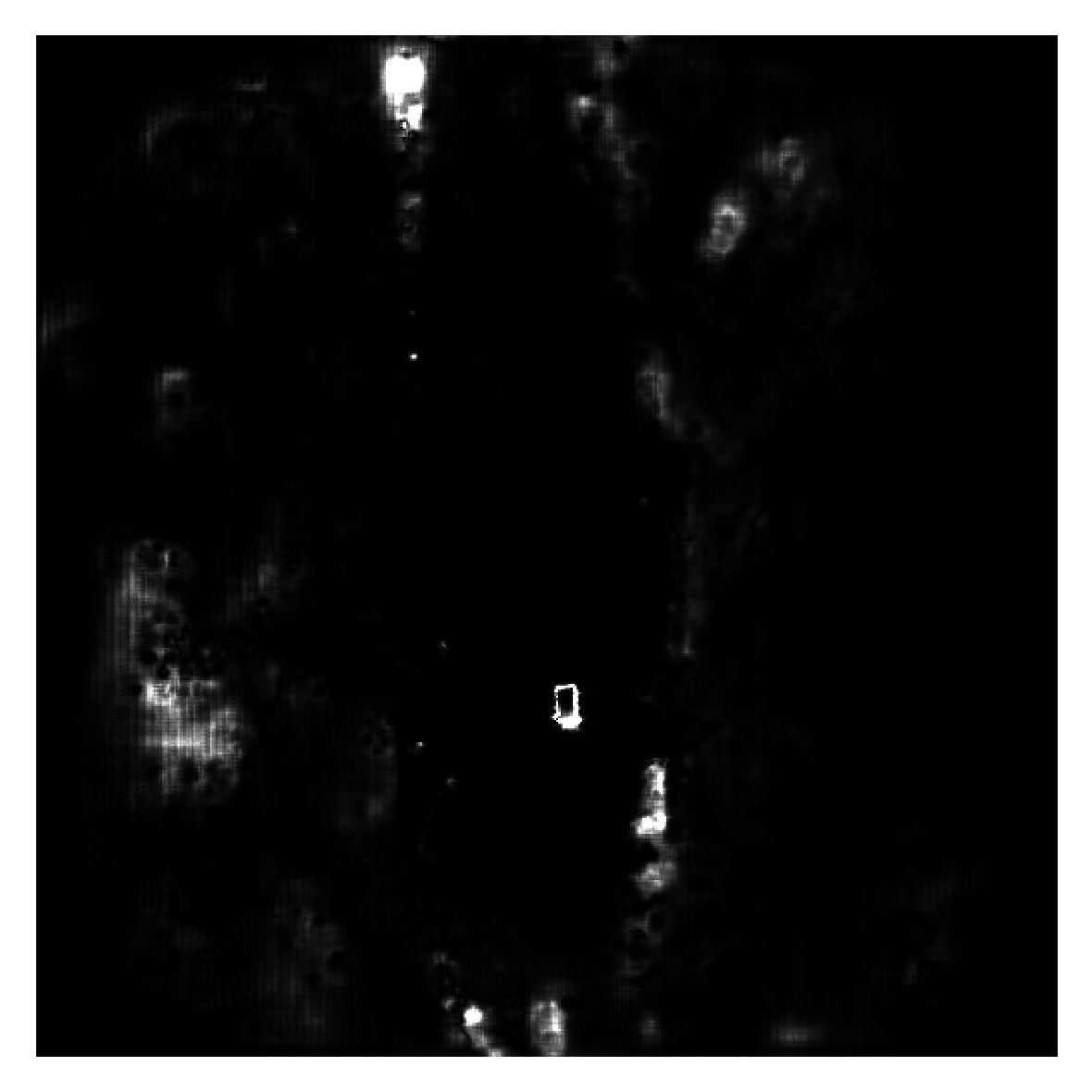}
     \caption{\scalebox{.7}{$M(\Phi'(X))_{\mathbf{obj}}$ - $M(\Phi(X))_{\mathbf{obj}}$}}
     \label{fig:img_benign}
     \end{subfigure}
    \end{minipage}
      \begin{minipage}{.24\linewidth}
     \begin{subfigure}{\linewidth}
     \centering
     \includegraphics[width=\linewidth]{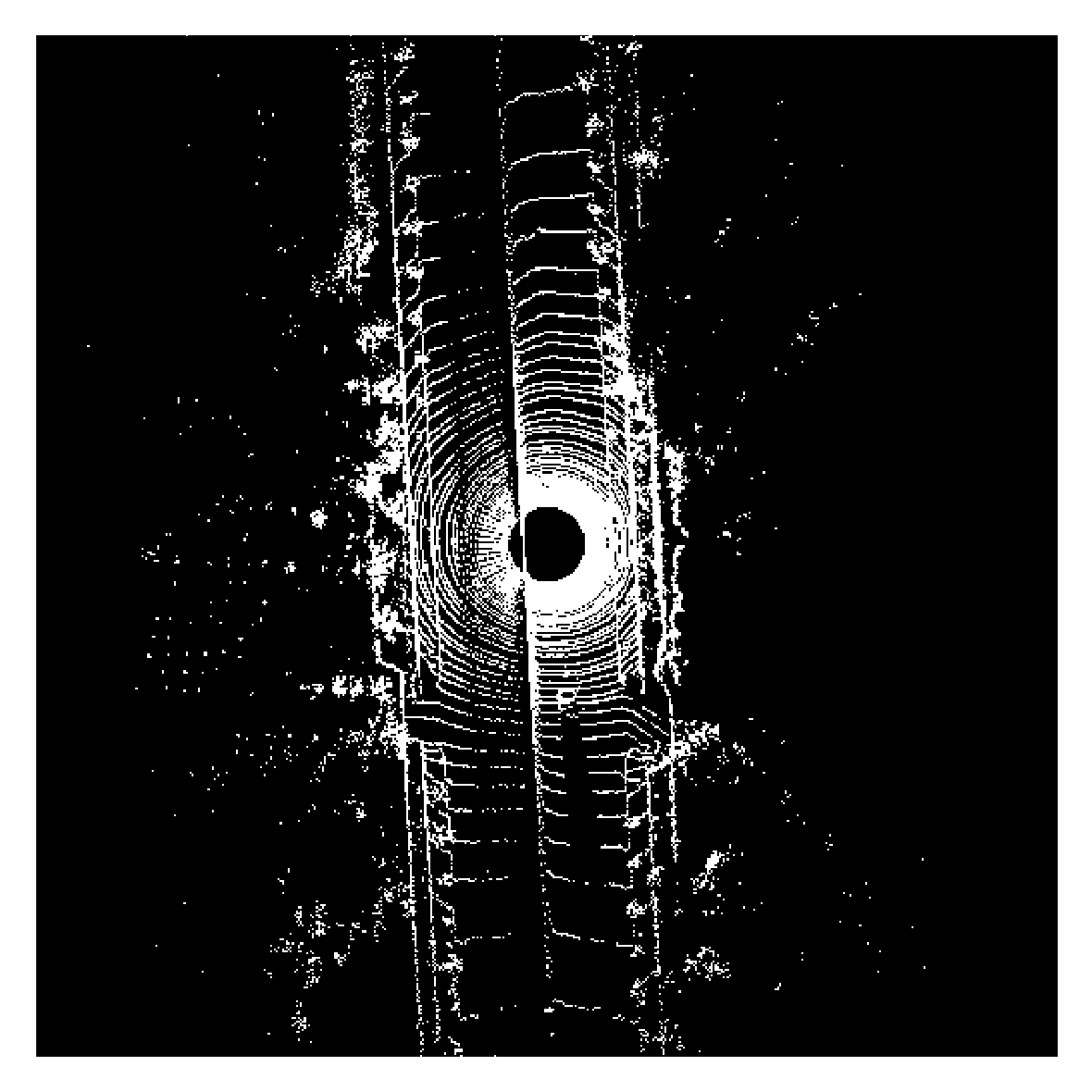}
     \caption{$\Phi''(X)_{\mathbf{count}}$ }
     \label{fig:img_benign}
     \end{subfigure}
    \end{minipage}
    \begin{minipage}{.24\linewidth}
     \begin{subfigure}{\linewidth}
     \centering
     \includegraphics[width=\linewidth]{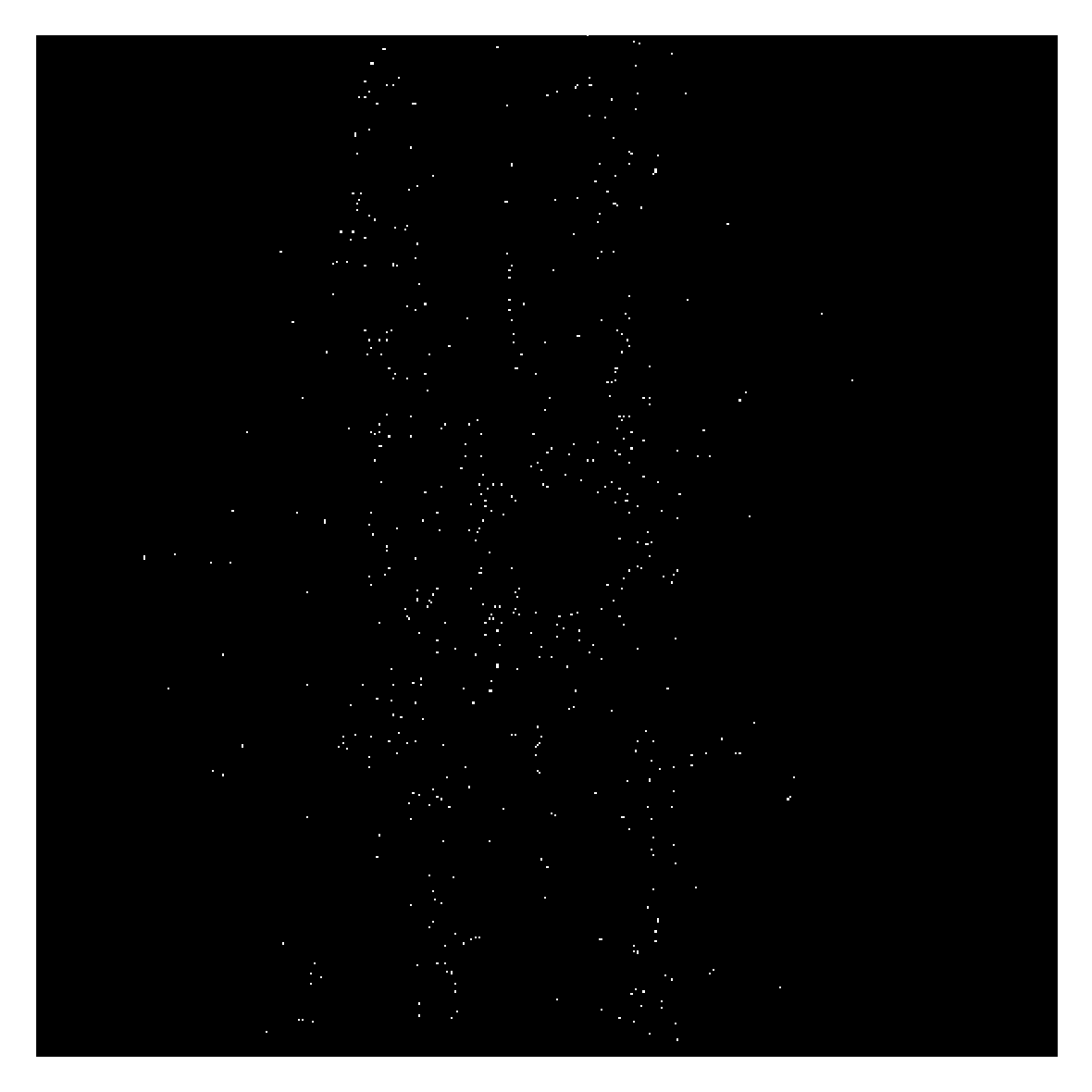}
     \caption{\scalebox{0.7}{$\Phi''(X)_{\mathbf{count}}$ - $\Phi(X)_{\mathbf{count}}$} }
     \label{fig:img_benign}
     \end{subfigure}
    \end{minipage}
     \begin{subfigure}{.240\linewidth}
     \centering
     \includegraphics[width=\linewidth]{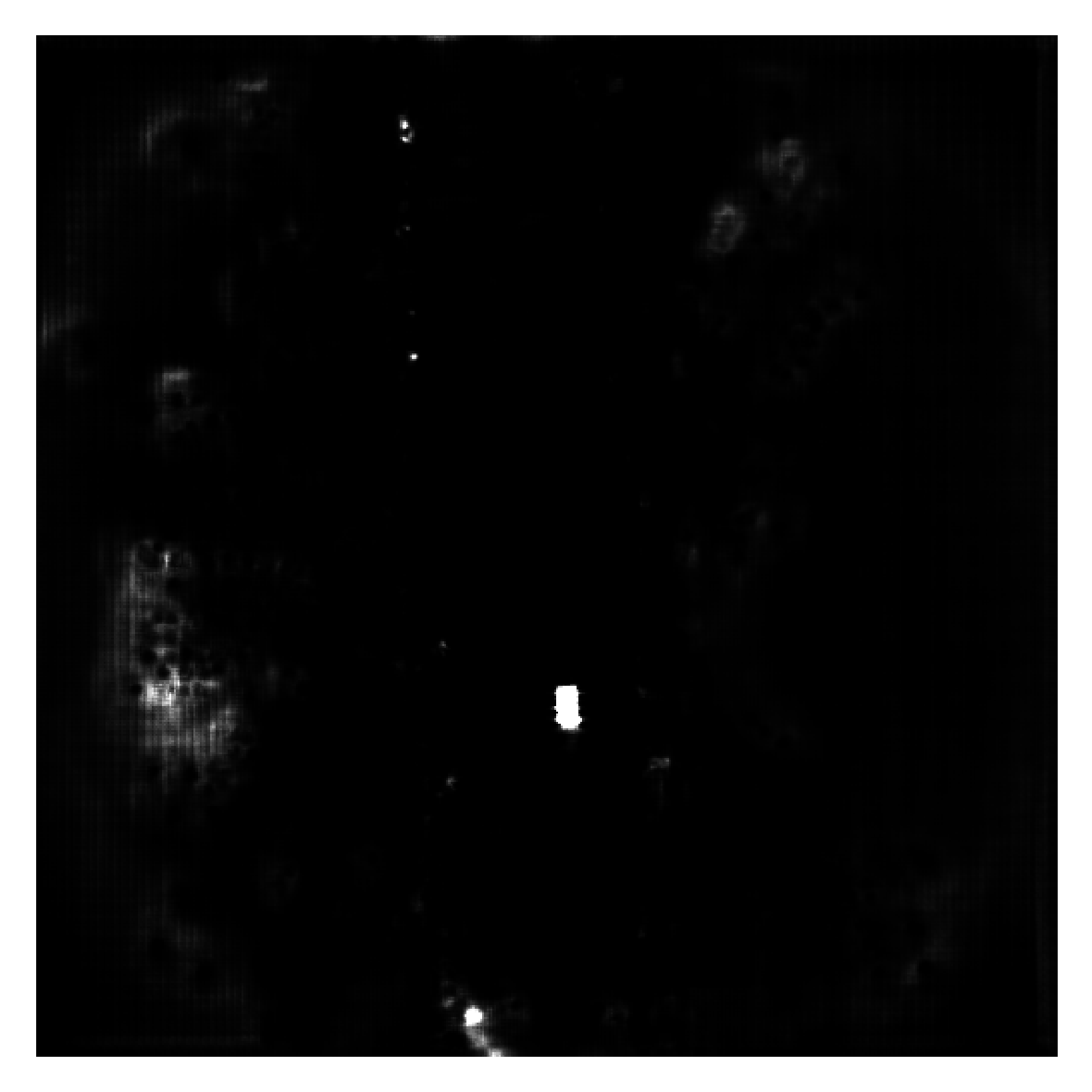}
     \caption{$M(\Phi''(X))_{\mathbf{obj}}$ }
     \label{fig:img_benign}
     \end{subfigure}
     \begin{subfigure}{.240\linewidth}
     \centering
     \includegraphics[width=\linewidth]{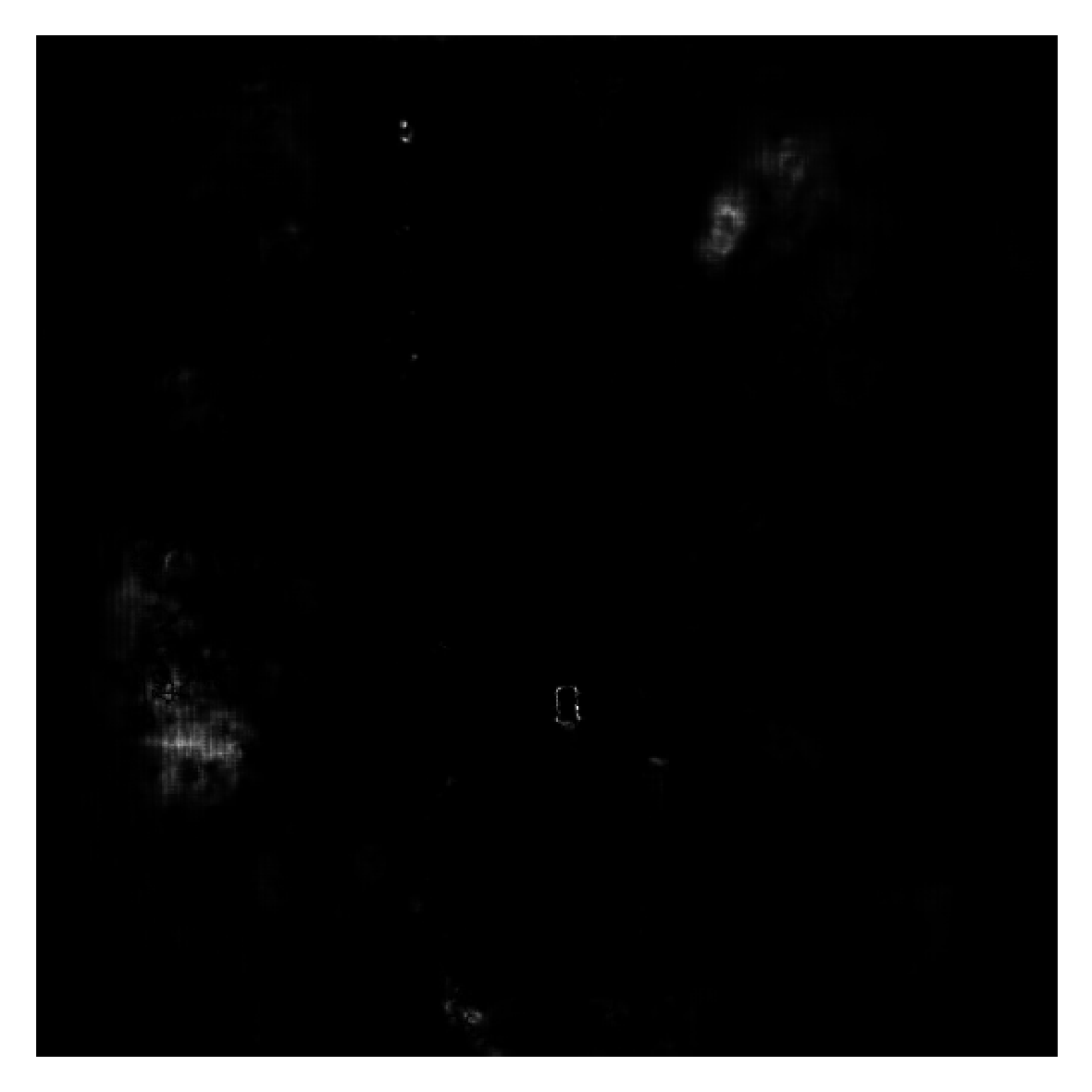}
     \caption{\scalebox{0.7}{$M(\Phi''(X))_{\mathbf{obj}}$-$M(\Phi(X))_{\mathbf{obj}}$}}
     \label{fig:img_benign}
     \end{subfigure}
    \caption{The performance of trilinear approximator and tanh approximator. The format $``\phi(\cdot)_{\mathbf{count}}''$  represents the 2D count feature calculated by trilinear approximator $\Phi'$; $M(\Phi'(X))_{\mathbf{obj}}$ represents visualization of  the ``objectiveness '' metric in the output of model $M$ using trilinear approximator with; $\Phi'(X)_{\mathbf{count}}$ - $\Phi(X)_{\mathbf{count}}$ represents the approximator 's error of $\Phi'$. The same notation for tanh approximiator $\Phi''$
    \label{fig:diff}
    }
\end{figure*}

Figure~\ref{fig:diff} visualizes the improvement of our tanh approximator $\Phi''$ compared to the trilinear approximator $\Phi'$ in terms of the \textbf{count} feature and the \textbf{objectness} metric. Given object $S$, $\Phi'(X)_{\mathbf{a}}$ represents the aggregated feature $a$ of the point cloud $X$. %
$M(\Phi'(X))_{\mathbf{a}}$ represents the model output with respect to metric $a$.  We observe that our approximator $\Phi'$ will introduce errors due to our approximation, which will finally leads to model output difference.
However, the error of the approximator has been largely decreased by using a more accurate approximator $\Phi''$. This reduces the error in model output, as can be seen in Figure~\ref{fig:diff}.

\section{Additional results}\label{sec:supp_additional_result}
\subsection{Changing label }
We  conduct experiments with  3 pristine meshes (cuba, sphere, tetrahedron) and set the target label to the other 4 labels except for the original label. The results are shown in \reftbl{changelabel-4mesh}, showing that our \StAdv has a high chance to trick the detector to output target labels, regardless of different pristine meshes that it starts from.

\begin{table}[tbh]
\centering
\caption{The attack success rate of the adversarial objects generating using \stAdv, starting from different types of pristine meshes. The target labels are the other four labels different from the original predictions.}
\label{tbl:changelabel-4mesh}
\centering
\begin{tabular}{ c c c c c c} 
\toprule
 & Cube &  Sphere & Tetrahedron & Cylinder & Overall\\
 \midrule
 Attack Success Rate &  75\% & 100\% & 75\% & 50\% & 75\% \\
\bottomrule
\end{tabular}
\label{tab:microwave}
\end{table}

\subsubsection{\stAdv on generating robust physical adversarial objects}
\cut{
\paragraph{\stAdv on different sizes}
The sizes of the pristine cube are set to $50$cm and $75$cm respectively.
We visualize the resulting meshes produced by \stAdv in Figure~\ref{fig:size}.

Note that both cubes have a reasonable shape with small number of vertices, but are able to fool the detector to miss them.
For the larger cube, it receives denser probing from the sensor, but is still able to achieve the goal.

Considering that the obstacles of $0.75$m and $1.5$m are as large as roadblocks in real life, this raises concerns that whether similar detection systems that are used in the real world are secure enough to defend such attacks.
}
In this subsection, we add additional results to evaluate the robustness of the generated objects against different positions and different angles. By doing so, it can provide insight on the performance of our adversarial object in real-world settings, before we 3D-print the object.

\begin{figure*}[htb]
    \centering
    \begin{minipage}{0.37\linewidth}
     \begin{subfigure}{\textwidth}
     \centering
     \includegraphics[width=\textwidth]{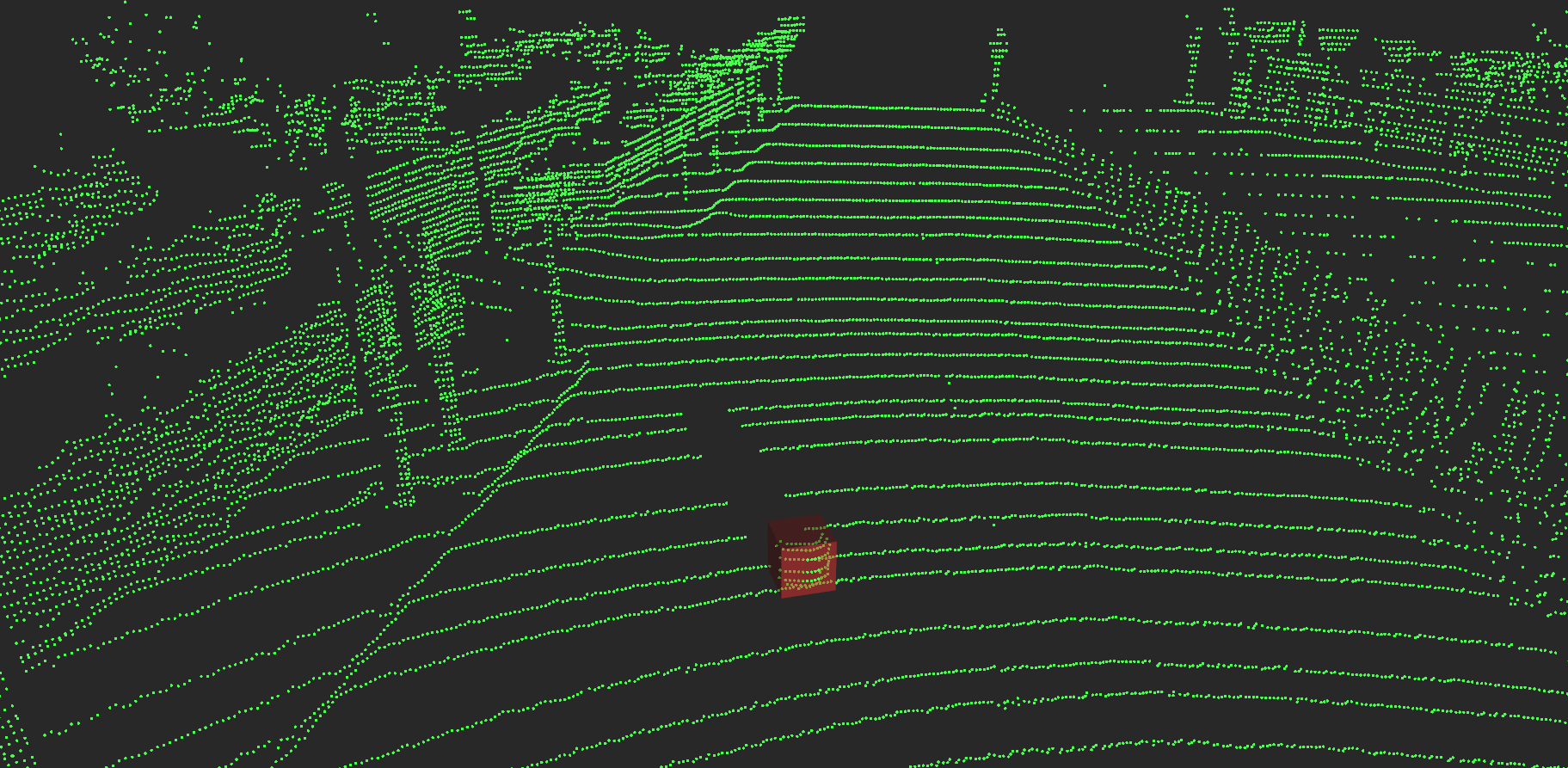}
     \caption{Benign frame}
     \label{fig:img_benign}
     \end{subfigure}
    \end{minipage}
    \begin{minipage}{.20\textwidth}
     \begin{subfigure}{\textwidth}
     \centering
     \includegraphics[width=\textwidth]{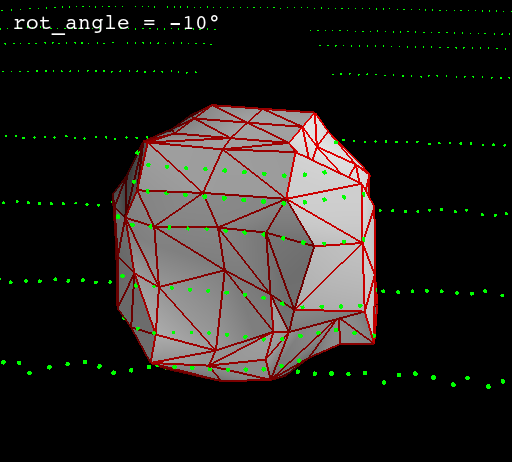}
     \caption{ Adv (-10\degree)}
     \label{fig:cross_entropy_benign}
     \end{subfigure}
    \end{minipage}
    \begin{minipage}{.20\textwidth}
     \begin{subfigure}{\textwidth}
     \centering
     \includegraphics[width=\textwidth]{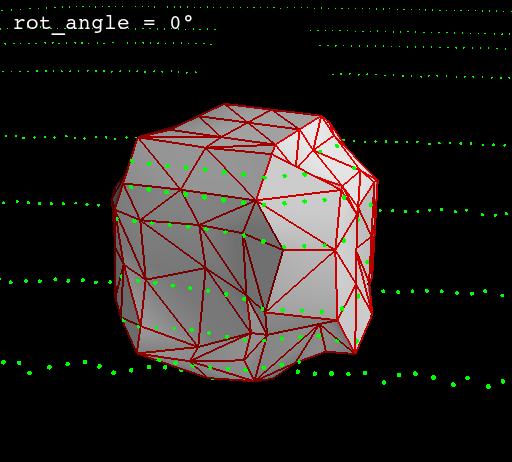}
     \caption{ Adv (0\degree)}
     \label{fig:cross_entropy_benign}
     \end{subfigure}
    \end{minipage}
    \begin{minipage}{.20\textwidth}
     \begin{subfigure}{\textwidth}
     \centering
     \includegraphics[width=\textwidth]{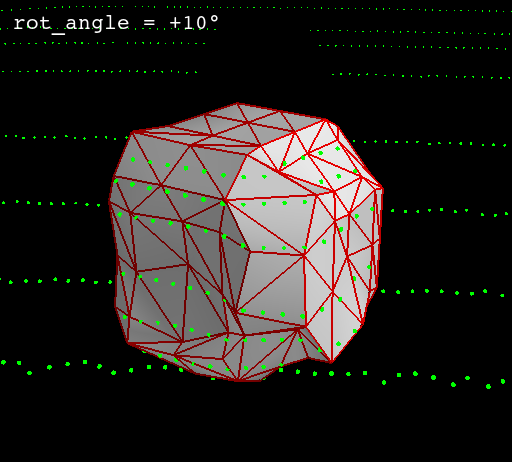}
     \caption{ Adv (10\degree)}
     \label{fig:cross_entropy_benign}
     \end{subfigure}
    \end{minipage}
    
    \caption{The visualization of adversarial object with different angles. In the benign frame (a), the system is able to detect the cube. When we replace the cube with our adversarial object, the system fails to detect the object at all three angles. We visualize the mesh along with the point clouds in a close-up view in (b), (c) and (d).}
    \label{fig:change}
\end{figure*}

\begin{table}[tb!]
\centering
\caption{Robust Adversarial Object against different angles. The original confidence is x. Our success rate is 100\%. (\checkmark represents no object detected)}
\centering
\begin{tabular}{ ccccccc } 
\toprule
\multicolumn{2}{c}{\shortstack{ Angle}}  &  -10\degree & -5\degree & 0\degree & 5\degree & 10\degree \\
\midrule
 \multirow{2}{*}{\shortstack{Objectness \\(Confid.)}}& Model &  \checkmark&   \checkmark&  \checkmark&   \checkmark&  \checkmark\\
  & Apollo &  \checkmark&   \checkmark &  \checkmark&   \checkmark &  \checkmark\\

\bottomrule
\end{tabular}
\label{tab:angle}
\end{table}
\paragraph{\stAdv against different angles}
We generate the adversarial objects by attacking for 9 angles simultaneously and evaluate the attack success rate among these angles. Our approach achieves 100\% attack success rate (Table~\ref{tab:angle}) %
both on our approximate model and the Apollo system.
This indicates that our designed differentiable proxy functions are accurate enough to transfer the adversarial behavior to Apollo. 
Figure~\ref{fig:change} shows qualitative results of the adversarial object from different close-up views. We can observe that the adversarial example is smooth and can be easily reconstructed in the real-world.
\begin{figure}[h]
    \centering
     \centering
     \includegraphics[width=\columnwidth{}]{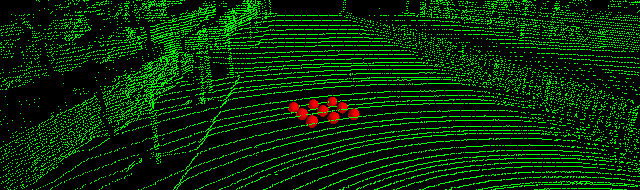}
    \caption{Our adversarial object can successfully attack the detection system, while placed at different positions. The red spheres mark the locations we place the adversarial object.}
    \label{fig:9loc}
\end{figure}

\begin{figure}[htb]
    \centering
    \includegraphics[width=\linewidth]{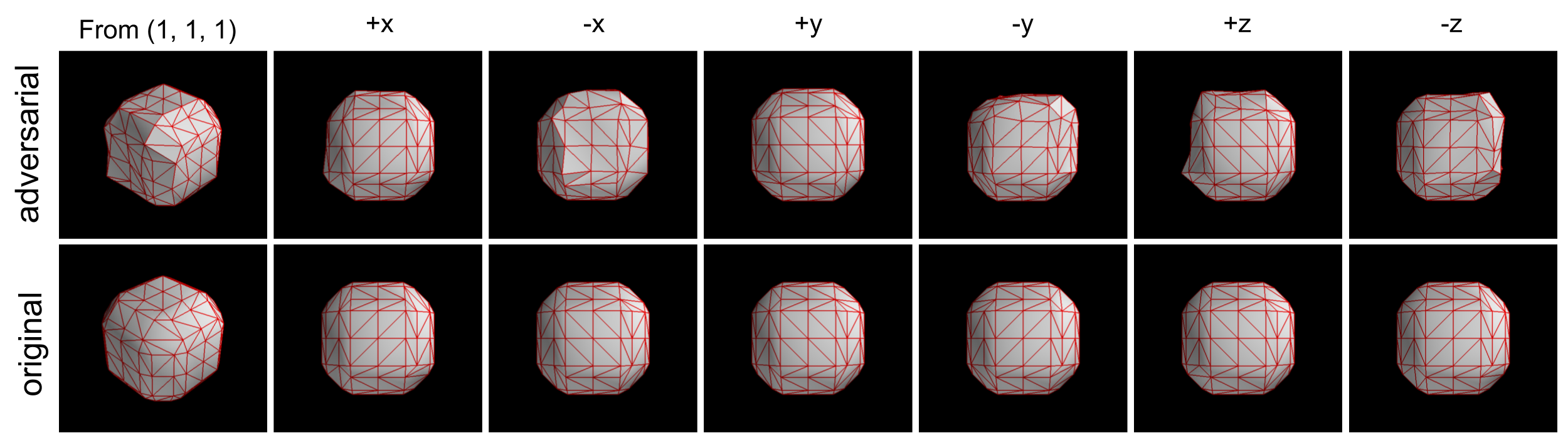}
    \caption{The optimized robust adversarial objects from 6 principal views and a particular view, compared with the original pristine object.}%
    \label{fig:vis_opps}
\end{figure}

\begin{table}[tb!]
\begin{small}
\centering
\caption{Robust Adversarial Object against different positions. The original object can be detected by Apollo. Our success rate is 100\%. (\checkmark represents no object detected)}
\resizebox{\linewidth}{!}{%
\begin{tabular}{c c c c c c c c  c c cc } 
\toprule
\multirow{2}{*}{\shortstack{Position}}  &  \multicolumn{2}{c}{Objectness (Confid.)} & \multirow{2}{*}{\shortstack{Position}}  &  \multicolumn{2}{c}{Objectness (Confid.)} & \multirow{2}{*}{\shortstack{Position}}  &  \multicolumn{2}{c}{Objectness (Confid.)}\\
& Ours  & Apollo & & Ours & Apollo & & Ours & Apollo \\
\toprule
(-50, -50)  & \checkmark &  \checkmark & (0, -50) &  \checkmark &  \checkmark  & (50, -50) & \checkmark &  \checkmark   \\
(-50, 0) &  \checkmark& \checkmark  & (0, 0 ) &  \checkmark &  \checkmark & (50, -50) & \checkmark &  \checkmark \\
(-50, 50) & \checkmark & \checkmark &  (0, 50) & \checkmark & \checkmark & (50, 50) & \checkmark &  \checkmark \\

\bottomrule
\end{tabular}
}
\label{tab:microwave}
\end{small}

\end{table}

\paragraph{\stAdv against different positions}
Similarly, we generate a single robust adversarial object against different positions simultaneously, as is shown in Figure~\ref{fig:9loc}.  We select 9 positions and use our algorithm to generate a universal robust adversarial example against different positions. 
Figure~\ref{fig:vis_opps} shows 7 views of the generated object from different angles, compared to the original object. This adversarial example is smooth from all views.
It shows that our approach is able to achieve the goal while keeping the shape plausible, so we can easily print the obect to perform physical attack.
Table~\ref{tab:microwave} show the detailed results of our adversarial object against these 9 positions: it can successfully attack the system among these 9 positions.

\cut{
\paragraph{\stAdv against different positions and different angles}
In this setting, we try to further enhance the robustness of our attacks by attack while varying both locations and angles, and test on unseen settings.
Table~\ref{tbl:all} shows the quantitative results.
An interesting phenomenon is that some results in the attack settings are worse than some results on unseen setting. This may imply that our adversarial objects are robust enough to generalize to unseen settings.
We can also safely conclude that our differentiable approximator is accurate enough to act as a proxy function such that produced adversarial objects can generalize.
}

\subsection{Physical experiments}

\begin{figure}[htb]
    \centering
    \begin{minipage}{0.62\linewidth}
     \begin{subfigure}{\linewidth}
     \centering
     \includegraphics[width=\linewidth]{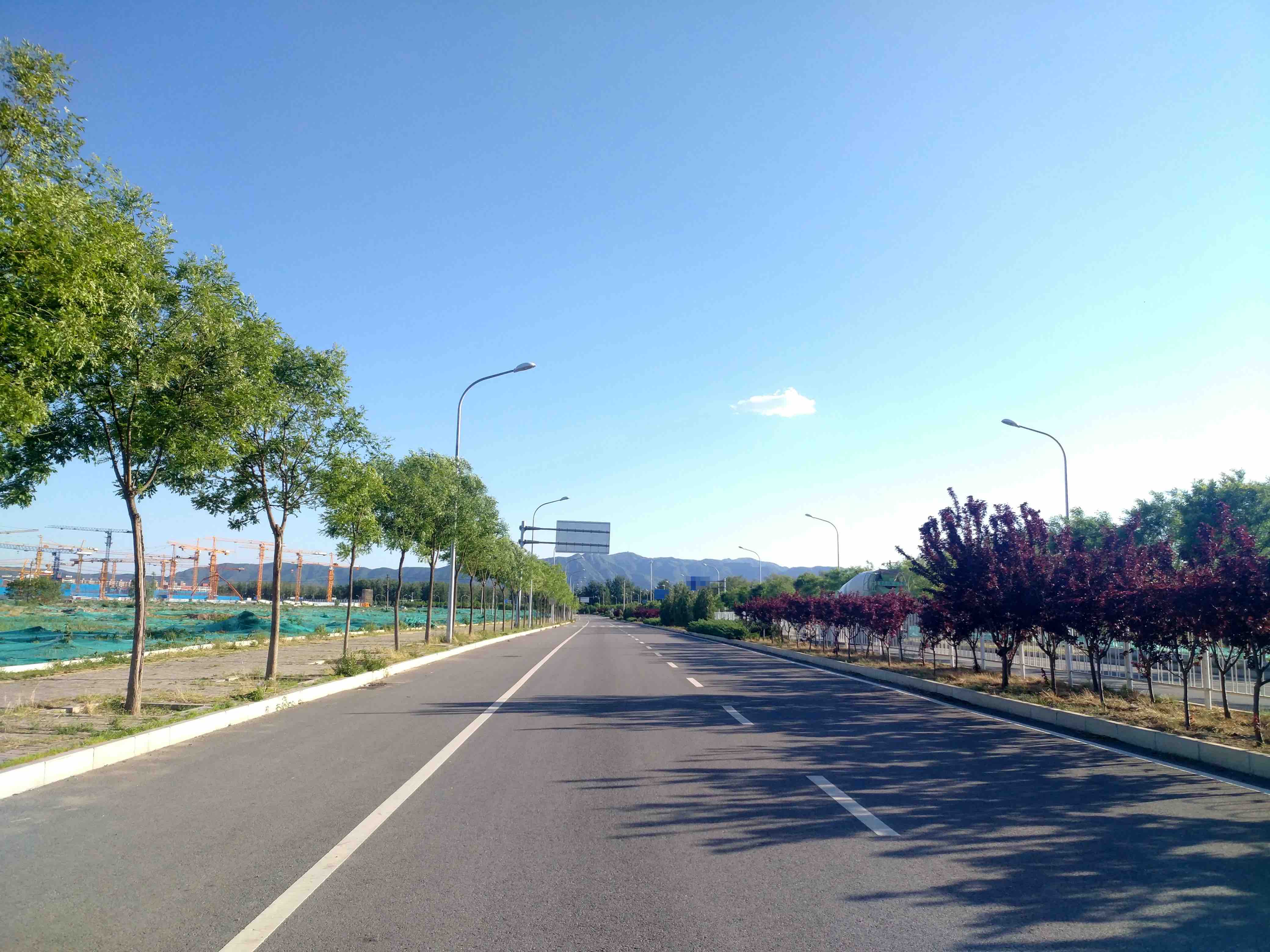}
     \caption{the road where we perform the physical experiment}
     \end{subfigure}
    \end{minipage}
    \begin{minipage}{.35\linewidth}
     \begin{subfigure}{\linewidth}
     \centering
     \includegraphics[width=\linewidth]{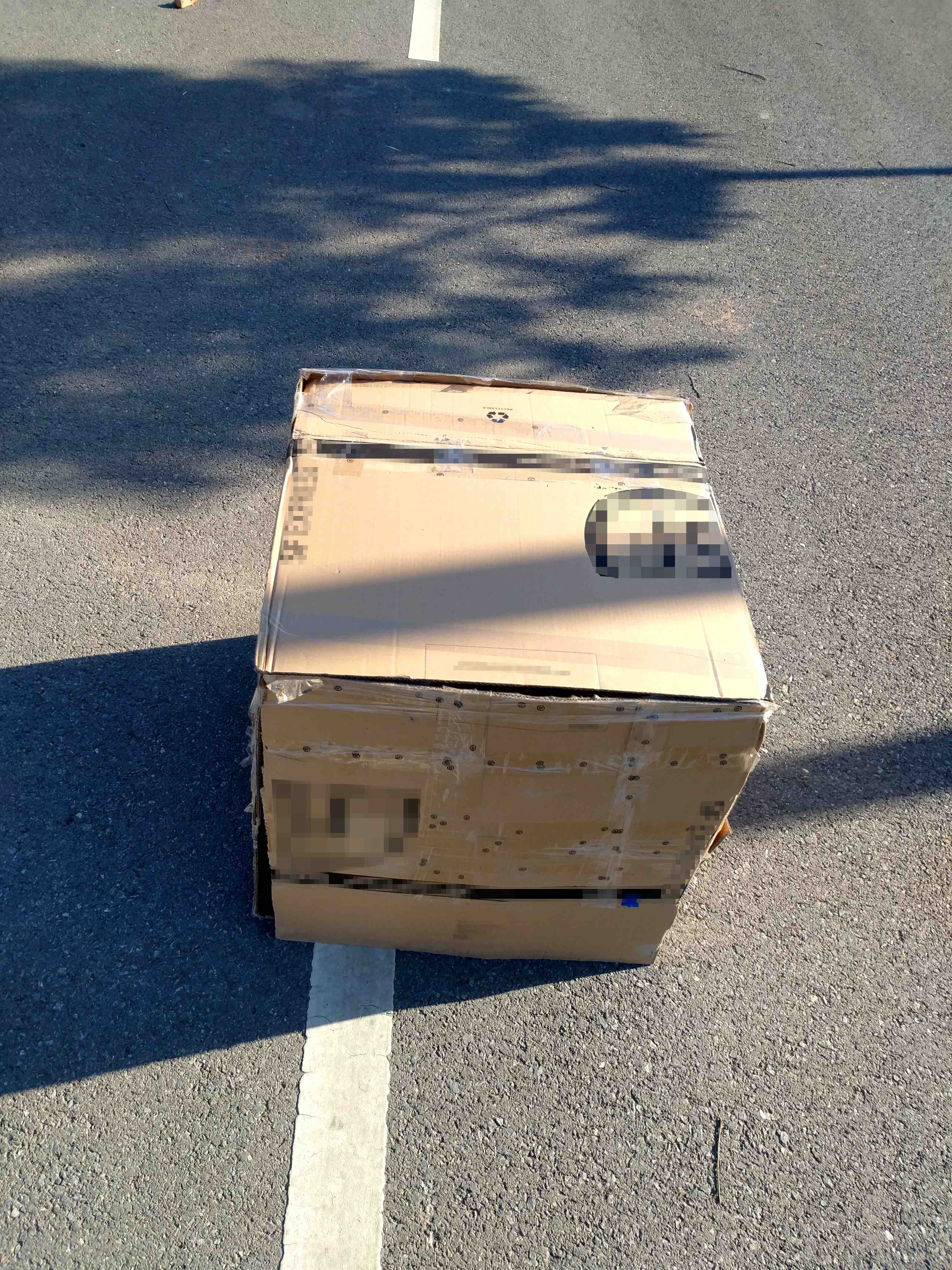}
     \caption{ the benign object for comparison}
     \end{subfigure}
    \end{minipage}
    \begin{minipage}{.62\linewidth}
     \begin{subfigure}{\linewidth}
     \centering
     \includegraphics[width=\linewidth]{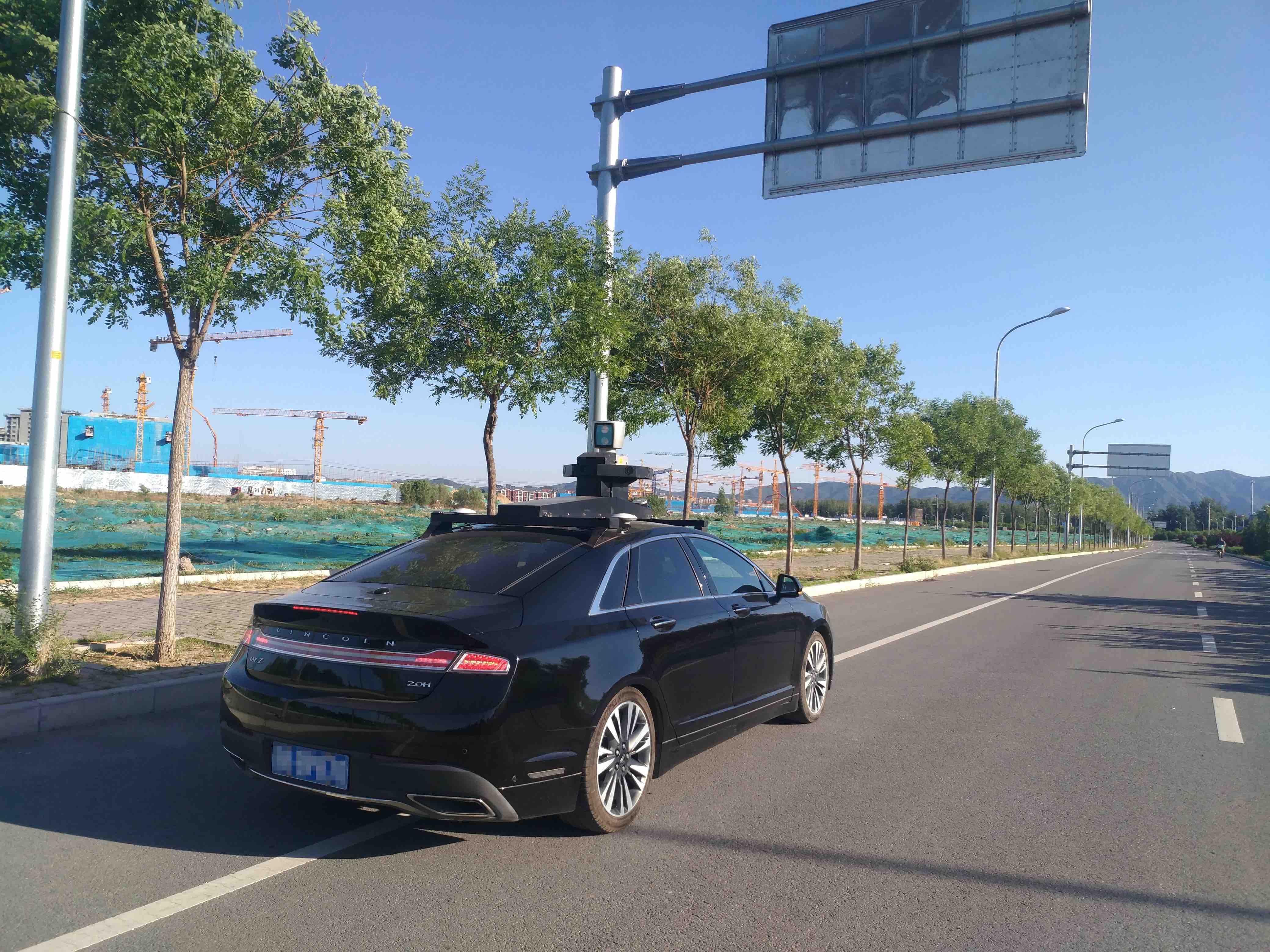}
     \caption{ the car used to collect the dashcam videos and the point clouds }
     \end{subfigure}
    \end{minipage}
    \begin{minipage}{.35\linewidth}
     \begin{subfigure}{\linewidth}
     \centering
     \includegraphics[width=\linewidth]{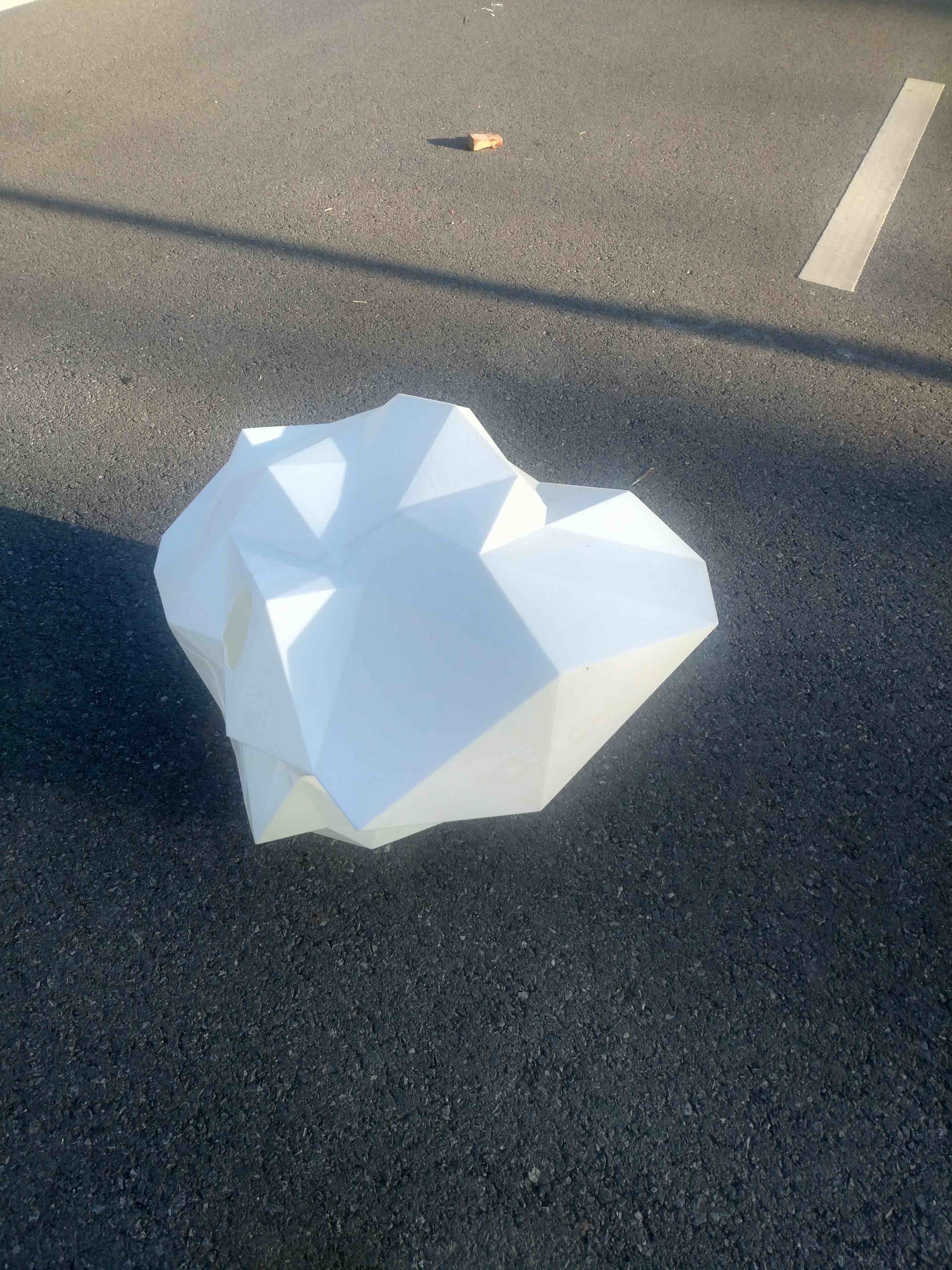}
     \caption{ our adversarial object}
     \end{subfigure}
    \end{minipage}
    \caption{Our physical experiment setting. We 3D-print the generated adversarial object at 1:1, and drive a car mounted with LiDAR and dashcams to collect the scanned point clouds and the reference videos. 
    }
    \label{fig:physical}
\end{figure}

We 3D-print our robust adversarial object at 1:1, and drive a real car mounted with LiDAR and dashcams. The adversarial object is put on the road, and a car drives by,
collecting scanned point clouds and the reference dashcam videos.
For comparison, we also put the benign object, which is a box of same size at the same location and follow the same protocol when collecting the point clouds.

\end{document}